%% file: main.tex
\documentclass[%
 reprint,
superscriptaddress,
 amsmath,amssymb,
 aps,
 pra,
]{revtex4-2}
\usepackage{graphicx}
\graphicspath{{figs/pdf/}{figs/tikz}}
\usepackage{dcolumn}
\usepackage{bm}
\usepackage{todonotes}

\usepackage{hyperref}

\usepackage{enumitem}
\usepackage{titlesec}
\usepackage{diagbox}
\usepackage{etoolbox}

\makeatletter
\renewcommand{\section}{%
  \@startsection{section}{1}{\z@}%
    {-3.5ex \@plus -1ex \@minus -.2ex}%
    {2.3ex \@plus.2ex}%
    {\normalfont\bfseries\scshape\centering}%
}
\makeatother

\titlespacing*{\subsection}
{0pt}{1.5ex plus 1ex minus .2ex}{1ex plus .2ex}
\titlespacing*{\subsubsection}
{0pt}{1.5ex plus 1ex minus .2ex}{1ex plus .2ex}
\titlespacing*{\paragraph}
  {\parindent}      
  {\medskipamount}  
  {1.5em}           

\usepackage{graphicx}
\usepackage{array, multirow}
\usepackage{physics}
\usepackage{ragged2e}

\usepackage{caption}
\usepackage{subcaption}
\DeclareCaptionFormat{justifcaption}{%
\begin{minipage}{\linewidth}\noindent\justifying\small #1#2#3\end{minipage}}
\DeclareCaptionFormat{centercaption}{%
\begin{minipage}{\linewidth}\noindent\Centering\small #1#2#3\end{minipage}}
\usepackage{tikz}
\usetikzlibrary{positioning,arrows.meta,calc}
\usepackage{standalone}

\usepackage{accents}
\newcommand{\thickbar}[1]{\accentset{\rule{.6em}{0.8pt}}{#1}}

\makeatletter
\renewcommand{\fnum@figure}{\textbf{Fig.~\thefigure}}  
\makeatother

\newcommand{\demxyz}{D_{XYZ}}
\newcommand{\demx}{D_{X}}
\newcommand{\demz}{D_{Z}}
\newcommand{\demxyzbar}{\thickbar{D}_{XYZ}}

\def\arxiversion{}

\ifdefined\arxiversion
    \newcommand{\heading}[1]{\section{#1}}
    \newcommand{\subheading}[1]{\subsection{#1}}
    \newcommand{\subsubheading}[1]{\subsubsection{#1}}
    \newcommand{\arxiv}[1]{#1}
    \newcommand{\natcomm}[1]{}
\else 
    \newcommand{\heading}[1]{\section*{#1}}
    \newcommand{\subheading}[1]{\subsection*{#1}}
    \newcommand{\subsubheading}[1]{\paragraph{#1:}}

    \newcommand{\arxiv}[1]{}
    \newcommand{\natcomm}[1]{#1}
    
    \makeatletter
    \patchcmd{\@startsection}
    {\@hangfrom{\csname #1name\endcsname \@seccntformat{#1}}}
    {\@hangfrom{\csname #1name\endcsname \@seccntformat{#1}}}
    {}{} 

    \renewcommand{\@seccntformat}[1]{%
      \ifcsname the#1\endcsname
       \csname the#1\endcsname\quad 
    \fi
    }
    \makeatother

    \preto{\subsection}{\setcounter{paragraph}{0}} 
\fi

\begin{document}

\title{\Large Decoding Correlated Errors in Quantum LDPC Codes}

\author{Arshpreet Singh Maan}
\affiliation{Aalto University, Espoo, Finland,\quad \{arshpreet.maan,\,alexandru.paler\}@aalto.fi}

\author{Francisco Miguel Garcia Herrero}%
\affiliation{Universidad Complutense de Madrid, Madrid, Spain,\quad francg18@ucm.es}

\author{Alexandru Paler}%
\affiliation{Aalto University, Espoo, Finland,\quad \{arshpreet.maan,\,alexandru.paler\}@aalto.fi}

\author{Valentin Savin}
\affiliation{Universit\'e Grenoble Alpes, CEA-L\'eti, Grenoble, France}
\affiliation{Quobly, Grenoble, France,\quad valentin.savin@quobly.io}


\begin{abstract}
We introduce a decoding framework for correlated errors in quantum LDPC codes under circuit-level noise.
The core of our approach is a graph augmentation and rewiring for inference (GARI) method, which modifies the correlated detector error model by eliminating 4-cycles involving $Y$-type errors, while preserving the equivalence of the decoding problem. We test our approach on the bivariate bicycle codes of distances $6$, $10$ and $12$. A normalized min-sum decoder with a hybrid serial-layered schedule is applied on the transformed graph, achieving high accuracy with low latency. Performance is further enhanced through ensemble decoding, where 24 randomized normalized min-sum decoders run in parallel on the transformed graph, yielding the highest reported accuracy (on par with XYZ-Relay-BP) with unprecedented speed for the tested codes under uniform depolarizing circuit-level noise. 
For the distance $12$ (gross) code, our approach yields a logical error rate of $(6.70\pm1.93) \times 10^{-9}$  at a practical physical error rate of $10^{-3}$. Furthermore, preliminary FPGA implementation results show that such high accuracy can be achieved in real time, with a per-round average decoding latency of 273\,ns and sub-microsecond latency in 99.99\% of the decoding instances. 
\end{abstract}

\maketitle

\heading{Introduction}
Quantum low-density parity-check (QLDPC) codes have recently emerged at the forefront of quantum error correction (QEC) research. Early interest was driven by the demonstration of constant-overhead fault tolerance using QLDPC codes~\cite{gottesman2013fault}, which spurred the search for code families that surpass the limitations of previous constructions defined by geometrically local stabilizers~\cite{bravyi2010tradeoffs}. These efforts have progressively advanced QLDPC code constructions~\cite{Pryadko_GenBicycleCodes, tillich_quantum_2014, breuckmann2020balanced, panteleev2021degenerate, panteleev2022quantumAlmostLinearMinD, pryadko_2bga, pacenti2025constructiondecodingquantummargulis}, resulting in asymptotically good families~\cite{panteleev_asymptotically_2021, leverrier2022_QTannerCodes, dinur_good_2022} (i.e., with non-vanishing rate and linear minimum distance), as well as practically implementable codes tailored to the constraints of specific quantum technologies~\cite{bravyi2024high, steffan2025tile}.

However, beyond code parameters and compatibility with quantum technology constraints, the merit of a QEC code also hinges on its practical decodability. At first glance, this may seem like good news for QLDPC codes, since decoding is what drove the remarkable success of their classical counterparts~\cite{savin2014ldpc}. 
Unfortunately, message-passing (MP) decoding algorithms -- a class of lightweight iterative algorithms with linear time complexity per iteration -- that are effective for classical codes fail in the quantum setting.
Indeed, for classical LDPC codes, both the theoretical and the practical significance of MP algorithms stems from their ability to perform optimal inference on cycle-free factor graphs (bipartite graphs representing the factorization of a joint probability distribution), where marginal probabilities and max-marginals can be computed via belief-propagation (BP) and min-sum (MS) algorithms, respectively~\natcomm{\cite{mezard2009information}}\arxiv{\cite{kschischang2001factor, mezard2009information}} 
(note that BP is also known as sum-product). In decoding terms, 
maximum a posteriori (MAP) decoding is realized via BP, while maximum likelihood (ML) decoding is realized via MS~\cite{wiberg1996codes}. And even though practical LDPC codes are defined by graphs with cycles, they locally resemble cycle-free graphs: their girth (the length of the shortest cycle) grows at the optimal rate $\Theta(\log n)$, where $n$ is the code length.  For QLDPC codes, the situation is markedly different: their low-weight stabilizers -- the very property that makes them suitable for fault-tolerance -- 
 typically lead to graphs with girth of $\Theta(1)$,  causing MP decoders to fail at producing sufficiently accurate inference~\cite{vasic2025quantum}. 

Short cycles (the shortest and most harmful being the 4-cycle) can lead to early error floors, where the decoder becomes trapped on certain error patterns~\cite{raveendran2021trapping,beni2025tesseractsearchbaseddecoderquantum},  
 causing a shallower -- potentially nearly flat -- decline in the logical error rate as the physical error rate decreases. This has motivated various approaches to enhance MP decoding performance, including heuristics to improve convergence and accuracy, post-processing techniques, and diversity or ensemble-based decoding methods (see related works, below). 
In contrast, our approach focuses on modifying the decoding graph rather than the MP decoding algorithm itself, while enabling ``vanilla'' MP decoders to perform their intended inference task more accurately.

Besides accuracy, the challenge of QLDPC decoding is further compounded by (1) circuit-level errors, captured by the detector error model, (2) the presence of correlations between $X$, $Y$, and $Z$ errors~\cite{beni2025tesseractsearchbaseddecoderquantum}, and (3) the requirement for real-time decoding, as timely error correction is essential to prevent backlog issues and maintain fault tolerance in quantum computation~\natcomm{\cite{RevModPhys.87.307, Battistel_2023}}\arxiv{\cite{RevModPhys.87.307, Battistel_2023,wang2024efficient}}. 
Our approach tackles these challenges and sets new benchmarks in decoding performance, achieving the highest accuracy with unprecedented speed for the bivariate bicycle (BB) codes of distance $6$ to $12$ from~\cite{bravyi2024high}. 

\arxiv{\subheading{Related Works}}
\natcomm{\medskip}
 Previous works have explored various strategies to improve MP decoding performance, particularly through combinations with post-processing techniques such as ordered-statistics decoding (OSD)~\cite{panteleev2021degenerate}.
This approach is usually referred to in the literature as BPOSD rather than MPOSD, even though BP decoding is rarely used and the best performance is typically achieved with normalized MS decoding~\cite{panteleev2021degenerate, ducrest2022stabilizer,roffe2020decoding}. Several subsequent works have continued to use the BP terminology in the more general sense of ``propagating belief messages'' -- that is, message passing -- regardless of whether the underlying inference would, in the cycle-free case, correspond to marginal probabilities or max-marginals.  In deference to the original works and to avoid any possible confusion, we adopt the same terminology as that used in the original papers whenever referring to previous literature. 

BPOSD quickly established itself as an almost universal solution~\cite{roffe2020decoding}, capable of decoding both code-capacity and circuit-level noise models, with an accuracy that remained the benchmark for some time. However, it relies on Gaussian elimination to produce the error estimate, which limits both parallelization and scalability for real-time applications, even on dedicated hardware~\cite{ducrest2024check, francisco_quantum_min_sum_fpga}. This has prompted the development of alternative approaches, which can be broadly classified into the following three categories: (1) Simplified post-processing methods, building upon or refining inversion-based or clustering approaches: e.g., localized statistics decoding~\cite{hillmann2024localized}, ordered Tanner forest decoding~\cite{demarti2024almost}, closed branch decoding~\cite{demarti2024closed}, ambiguity clustering~\cite{wolanski2024ambiguity}; (2) Heuristic methods for improving the MP decoding performance by modifying the decoding rules: e.g., guided decimation guessing~\cite{gong2024lowlatencyiterativedecodingqldpc}, collaborative check removal~\cite{bhattacharyya2025decoding}, symmetry-breaking techniques~\cite{yin2024symbreak}, refined or memory BP~\cite{kuo_refined_2020,kuo2022exploitingdegeneracyNature}, MP decoders with past influence~\cite{chytas2025enhancedMSIterdynamics}; (3) Diversity or ensemble-decoding techniques, which combine multiple decoders to improve overall performance and mitigate individual decoder failures: e.g., automorphism ensemble decoding~\cite{koutsioumpas2025automorphismensembledecodingquantum} or Relay-BP~\cite{muller2025improvedbeliefpropagationsufficient}.  Some approaches span multiple categories, e.g., stabilizer inactivation~\cite{ducrest2022stabilizer},  check-agnosia~\cite{ducrest2024check}, or speculative decoding~\cite{wang2025fullyparallelizedbpdecoding}, which are simplified post-processing methods, where the post-processing is based on ensemble MP-decoding.  

These methods still face major limitations, including high resource demands and reduced performance when decoding correlated errors (i.e., correlations between $X$, $Y$, and $Z$ errors). For instance, Ref.~\cite{beni2025tesseractsearchbaseddecoderquantum} shows that BPOSD is limited by its inability to decode correlated errors, as BP struggles with short cycles in the correlated detector error model, providing misleading information to the OSD post-processor. The XYZ-Relay-BP decoder~\cite{muller2025improvedbeliefpropagationsufficient} efficiently handles correlated errors,  significantly outperforming BPOSD, by employing an ensemble decoding approach that uses a large number of BP decoders sequentially. Nevertheless, while the approach can be regarded as a lightweight MP decoder, its sequential nature may result in increased latency.

 Graph-search-based decoders such as Tesseract~\cite{beni2025tesseractsearchbaseddecoderquantum} and decision-tree decoders~\cite{ott2025decisiontreedecodersgeneralquantum} have emerged as promising solutions, providing the most accurate decoding of correlated errors reported so far. However, this comes at the cost of increased complexity, which may preclude their use in large-scale applications.

\begin{figure}[t]
\includegraphics[width=.49\textwidth]{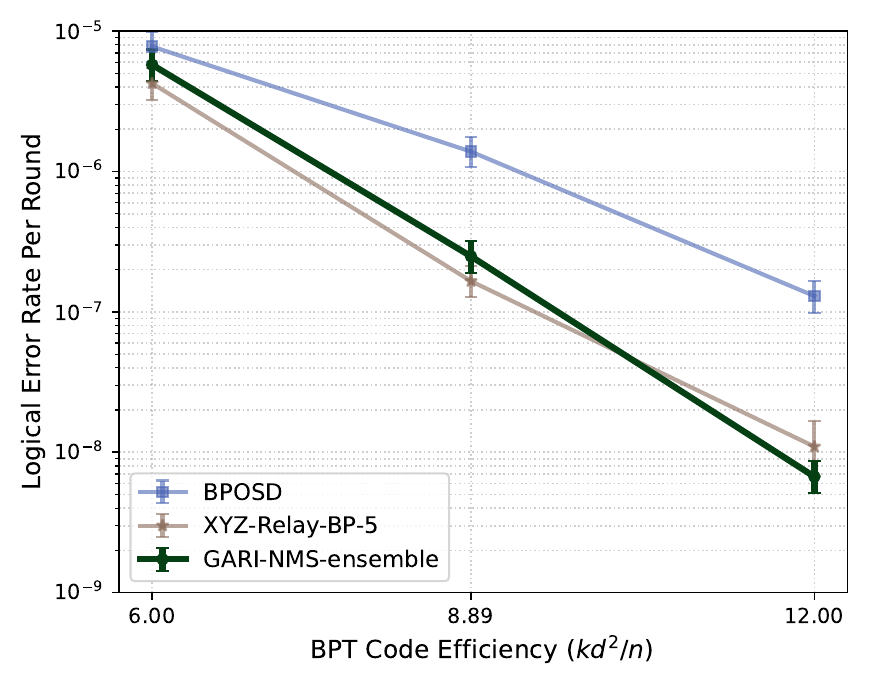}
\caption{Logical error rate performance of our method (GARI-NMS-ensemble), compared with BPOSD and XYZ-Relay-BP-5~\cite{muller2025improvedbeliefpropagationsufficient}. We consider BB codes with $[[n,k,d]]$ parameters equal to $[[72,12,6]]$, $[[90,8,10]]$ and $[[144,12,12]]$, under uniform depolarizing circuit level noise with a physical error rate of 0.001. We plot logical error rates against the \emph{BPT efficiency} of the code, defined as $kd^2/n$, which accounts for the non-uniform variation of the $n,k,d$ parameters, and is inspired by the Bravyi-Poulin-Terhal bound~\cite{bravyi2010tradeoffs}. Error bars indicate 99\% confidence intervals.
For BPOSD, we report the best achievable performance obtained by fine-tuning both the MP decoder and its underlying message schedule.  The XYZ-Relay-BP-5 decoder uses  301 DMem-BP decoders in series with up to $300\times 60 + 80  =$ 18080 decoding iterations, and memory strength parameters taken from~\cite{muller2025improvedbeliefpropagationsufficient}. It searches for five different solutions and returns the one with the lowest weight.
The GARI-NMS-ensemble decoder employs 24 NMS decoders in parallel for BB codes with distances 10 and 12, with each decoder performing up to 400 iterations.  We adopt a minimum-latency approach, in which decoding stops as soon as one decoder in the ensemble converges, and the corresponding solution is then returned. For distance 6, only a single GARI-NMS decoder is used,  as employing multiple decoders does not further improve performance.}
\label{fig:logical-error-rate-comp}
\end{figure}

\arxiv{\subheading{Contributions}}
\natcomm{\medskip}
In this work, we focus on MP decoding of correlated errors under the circuit-level noise model. The starting observation is that the inefficiency of MP decoding in performing the intended inference stems from the structural properties of the underlying decoding graph. Accordingly, our contributions center on modifying the decoding graph to improve its suitability for MP decoding, and on exploiting the new structure via an expert selection of MP decoding parameters.

Considering the correlated detector error model, the decoding graph is a bipartite graph involving error nodes and check nodes. Error nodes correspond to $X$, $Y$, or $Z$ errors propagating from different locations in the syndrome measurement circuit (possibly merging multiple locations if they produce the same syndrome), while check nodes correspond to either $X$ or $Z$ detectors (we assume CSS codes). The core of our approach is a graph augmentation and rewiring technique that transforms the original detector error model graph into a new graph representing an equivalent decoding problem, while eliminating 4-cycles passing through $Y$-type error nodes. Eliminating 4-cycles is intended to facilitate MP inference; thus, we refer to the proposed graph transformation as  \emph{graph augmentation and rewiring for inference} (abrv. GARI). 

We then perform MP decoding on the new graph obtained after applying GARI. Our MP decoding consists of a normalized MS (NMS) algorithm with a hybrid serial-layered message schedule, chosen to optimize the decoding accuracy while also reducing the overall latency. For the tested BB codes, this results in a logical error rate matching that of the best-performing BPOSD, with OSD of order-zero. 

To further improve the decoding performance, we employ an ensemble decoding approach, running multiple MP decoders in parallel on our augmented and rewired decoding graph. To this end, we leverage the randomization in the serial part of the hybrid serial-layered schedule, using a modest-sized ensemble of MP decoders with different randomization seeds. 
This approach achieves the highest decoding accuracy reported to date, on par with  XYZ-Relay-BP-5 (see Fig.~\ref{fig:logical-error-rate-comp}). 
Moreover, preliminary hardware implementation results for the [[144,12,12]] BB code show that each decoder of the ensemble can run in real-time on a single VU19P FPGA device. For instance, at a physical error rate of $10^{-3}$, the average decoding latency is 273\,ns, and in 99.99\% of instances the decoding completes within 1\,$\mu$s per round.

\arxiv{\heading{Methods}}
\natcomm{\heading{Results}}

\subheading{Graph Augmentation and Rewiring for Inference}
\arxiv{\label{sec:gari}}
A decoding problem can typically be formulated as a linear system, where the decoder's task is to identify the most likely solution given the prior probability distributions of the errors (variables of the system). The linear system is defined by a matrix, which we call the decoding matrix. For MP decoders, the decoding matrix can be conveniently represented as a decoding graph: a bipartite graph whose adjacency matrix is the decoding matrix, with vertices referred to as error nodes and check nodes, corresponding to the columns and rows of the matrix, respectively (we use the term error node instead of the variable node or data node terminology commonly used in the literature, to explicitly indicate that these nodes represent Pauli errors). We describe the GARI method in terms of both the decoding matrix\arxiv{ (Section~\ref{sec:matrix}} and the decoding graph\arxiv{ (Section~\ref{sec:graph})}. The first description, based on the decoding matrix, focuses on the correlated detector error model and provides a simple linear-algebra procedure to construct the GARI matrix. The second description, based on the decoding graph, applies to more general graph models, and can be used for further optimization and refinement of our approach.

\begin{figure*}
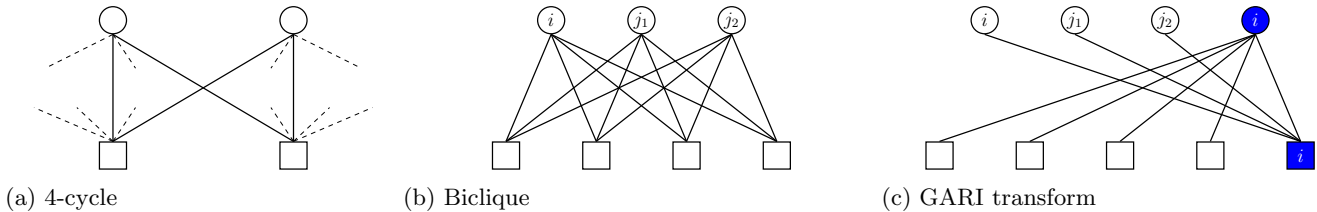

\centering
\begin{subfigure}[b]{0.29\textwidth}
    \centering
    \includestandalone[scale=.6]{4cycle}  
    \caption{4-cycle}
    \label{fig:4-cycle}
\end{subfigure}%
\hfill%
\begin{subfigure}[b]{0.35\textwidth}
    \centering
    \includestandalone[scale=.6]{biclique}  
    \caption{Biclique}
    \label{fig:biclique}
\end{subfigure}%
\hfill%
\begin{subfigure}[b]{0.35\textwidth}
    \centering
    \includestandalone[scale=.6]{gari}  
    \caption{GARI transform}
    \label{fig:gari}
\end{subfigure}
\caption{Example of GARI transformation. (a)~A 4-cycle in a decoding graph: error and check nodes are depicted as circles and squares, respectively; edges forming the 4-cycle are shown in bold, other incident edges are shown as dashed.  (b)~A biclique in the decoding graph; only edges participating in the biclique are shown. The MP inference is not exact, due to the presence of the 4-cycles. (c)~The GARI transform eliminates all 4-cycles forming the biclique, by performing a variable change that introduces a new error node and a new check node in the graph (depicted in blue), as explained in the main text.  If the eliminated 4-cycles were the only cycles in the bipartite graph, the new graph would be cycle-free, and the MP inference would be exact. The error and check node labels in (b) and (c) are included to illustrate the connection with the decoding-matrix perspective\arxiv{ in Section~\ref{sec:matrix}} and should be disregarded unless specifically referenced in the main text.}
    \label{fig:improved-MP-inference}
\end{figure*}

\subsubheading{The Decoding Matrix Perspective} 
\arxiv{\label{sec:matrix}}
We assume Calderbank–Shor–Steane (CSS) codes. Thus, in the correlated detector error model~\cite{derks2024designingfaulttolerantcircuitsusing,higgott2024practical}, the decoding matrix is a binary matrix that takes the following form:
\begin{equation}
\label{eq:demxyz-matrix}
\renewcommand{\arraystretch}{1.5}
    \begin{array}{cc@{\;\;}|@{\;\;}c@{\;\;}|@{\;\;}ccc}
    & \multicolumn{1}{c@{\;\;}@{\;\;}}{e_Z} & \multicolumn{1}{c@{\;\;}@{\;\;}}{e_X} & \multicolumn{1}{c@{\;\;}}{e_Y} & & \\
    \multirow{2}{*}{$\demxyz = $\Bigg(} & 
    \demx & 0 & \demx' & 
    \multirow{2}{*}{\Bigg)} & s_X \\
    \cline{2-4}
    & 0 & \demz & \demz' & & s_Z
    \end{array}
\end{equation}
The matrix is partitioned into three block columns: the left block corresponds to $Z$ errors, the middle block to $X$ errors, and the right block to $Y$ errors. The corresponding errors are denoted by $e_Z$, $e_X$, and $e_Y$. 
Similarly, the rows are divided into two blocks: the top block corresponds to $X$-type detectors, and the bottom block to $Z$-type detectors. The corresponding syndromes are denoted by $s_X$ and $s_Z$.

For given syndromes $s_X$ and $s_Z$, the role of the decoder is to estimate the most likely error vector $(e_Z, e_X,e_Y)$, such that  (we always treat vectors as column vectors in equations; however, error vectors $e_Z$, $e_X$ and $e_Y$ should be interpreted as row vectors when used to label the columns of the decoding matrix): 
\begin{equation}
\label{eq:H-system}
    \demxyz (e_Z, e_X, e_Y) = (s_X, s_Z).
\end{equation}
Note that matrices $\demx$ and $\demz$ are those obtained when considering only one type of detectors, that is, either $X$ or $Z$ detectors, which detect  $Z$ or $X$ errors, respectively. The notation $\demxyz$ for the correlated detector error model is somewhat abusive, since there are no $Y$-type detectors.  Since our focus is on correlated errors, we will assume that any circuit location where a $Y$ error may occur can also experience $X$ and $Z$ errors. Therefore,  matrices $\demx$ and $\demx'$ satisfy the following properties, and analogous properties hold for $\demz$ and $\demz'$ as well: 
\begin{itemize}[topsep=0pt, partopsep=0pt, itemsep=0pt]
    \item  $\demx$ has no repeating columns,
    \item every column of $\demx'$ is equal to a column of $\demx$; however, $\demx'$ may contain repeated columns, meaning the same column of $\demx$ may appear multiple times in $\demx'$.
\end{itemize}
The fact that $\demx$ has no repeating columns follows directly from the definition of the detector error model:  $Z$ errors occurring at different locations but producing the same syndrome are merged into a single column. 
For the second property, we observe that a $Y$-error and a $Z$-error  occurring at the same location produce the same $X$-syndrome.  This explains why every column of $\demx'$ is equal to a column of $\demx$. However, $\demx'$ may contain repeated columns, corresponding to $Y$-errors at different locations that produce the same $X$-syndrome but different $Z$-syndromes. In this case, the corresponding columns are not merged in $[\demx'; \demz']$, resulting in column repetitions in  $D'_X$ (where we use the semicolon to denote vertical stacking of matrices).   
These properties imply the existence of matrices $U$ and $V$, such that $\demx' = \demx U$ and $\demz' = \demz V$, which can be explicitly determined as follows:
\begin{itemize}[topsep=0pt, partopsep=0pt, itemsep=0pt]
    \item  each column of $U$ has exactly one $1$, and
    \item the $i$th row of $U$ has $1$'s in columns $j_1,\dots,j_k$ if and only if the $i$th column of $\demx$ equals the columns  $j_1,\dots,j_k$ in $\demx'$.
    \item Similar statements hold for $V$.
\end{itemize}

We use matrices $U$ and $V$ to perform a change of variables in Eq.~\eqref{eq:H-system}. Precisely, we define
\begin{equation}
\label{eq:bar-errors}
    \bar{e}_Z = e_Z + U e_Y \quad \text{ and } \quad \bar{e}_X = e_X + V e_Y,
\end{equation}
so that Eq.~\eqref{eq:H-system} rewrites as 
\begin{equation}
\label{eq:H-system-rewrite}
    \demx \bar{e}_Z = s_X \quad \text{ and } \quad \demz \bar{e}_X = s_Z.
\end{equation}
Independent decoders operating on $D_X$ and $D_Z$ would provide independent estimations of $\bar{e}_X$ and $\bar{e}_Z$ satisfying Eq.~\eqref{eq:H-system-rewrite}. To account for correlations between $X$, $Y$, and $Z$ errors, a decoder must jointly estimate $\bar{e}_X$ and $\bar{e}_Z$ satisfying Eq.~\eqref{eq:H-system-rewrite} along with the variable change in Eq.~\eqref{eq:bar-errors}. Precisely, this corresponds to defining a new decoding matrix
\begin{equation}
\label{eq:gari-matrix}
\renewcommand{\arraystretch}{1.5}
    \begin{array}{cc@{\;\;}|@{\;\;}c@{\;\;}|@{\;\;}c@{\;\;}|@{\;\;}c@{\;\;}|@{\;\;}ccc}
    & \multicolumn{1}{c@{\;\;}@{\;\;}}{e_Z} & \multicolumn{1}{c@{\;\;}@{\;\;}}{e_X} & \multicolumn{1}{c@{\;\;}@{\;\;}}{e_Y} & \multicolumn{1}{c@{\;\;}@{\;\;}}{\bar{e}_Z} & \multicolumn{1}{c@{\;\;}}{\bar{e}_X} & & \\
    \multirow{4}{*}{$\demxyzbar = \left(\rule{0pt}{37pt}\right.$} & 
    0 & 0 & 0 & \demx & 0 & 
    \multirow{4}{*}{$\left.\rule{0pt}{37pt}\right)$} & s_X \\
    \cline{2-6}
    & 0 & 0 & 0 & 0 & \demz &  & s_Z \\
     \cline{2-6}
    & I & 0 & U & I & 0 &  & 0 \\
     \cline{2-6}
    & 0 & I & V & 0 & I &  & 0 
    \end{array}
\end{equation}
so that Eq.~\eqref{eq:H-system-rewrite} and Eq.~\eqref{eq:bar-errors} can be rewritten as
\begin{equation}
\label{eq:H-system-gari}
    \demxyzbar (e_Z, e_X, e_Y, \bar{e}_Z, \bar{e}_X) = (s_X, s_Z, 0, 0).
\end{equation}
Note that Eq.~\eqref{eq:H-system-gari} depends on an augmented variable set, consisting of $(e_Z, e_X, e_Y, \bar{e}_Z, \bar{e}_X)$. We refer to $\demxyzbar$, which defines an equivalent decoding problem, as the GARI matrix.  Importantly, a GARI decoder -- i.e., one using the decoding matrix $\demxyzbar$ -- produces an estimate of the error vector $(e_Z, e_X, e_Y, \bar{e}_Z, \bar{e}_X)$. In practice, however, we are typically interested in only one type of error, depending on the type of the logical observables protected by the decoder. Assuming we protect logical-$Z$ observables, we are only interested in decoding $X$ errors,  including contributions from $Y$ errors, that is $\bar{e}_X$. This is precisely the error that would be estimated using the $Z$-type detector error model alone; however, in that case, the decoder cannot exploit correlations between the Pauli errors.

\begin{table*}    
    \centering
    \caption{Size, average row weight  ($\bar{w}_r$, a.k.a. average check degree), and number of 4-cycles in  $\demx$, $\demz$, $\demxyz$ and $\protect\demxyzbar$ for BB codes with $[[n,k,d]]$ parameters indicated in the left column. Detector error model matrices are generated using the Stim circuits from~\cite{gong2024lowlatencyiterativedecodingqldpc}, where, for a code of distance $d$, the syndrome measurement is repeated $d$ times, followed by a final measurement of the data qubits in the $Z$ basis. The number of 4-cycles removed by GARI corresponds to the difference between the number of 4-cycles in $\demxyz$ and the combined total in $\demx$ and $\demz$ (e.g., $11584296 - (47232+53280) = 11483784$ for the distance 12 code). For $\protect\demxyzbar$ we report the size and average row weight of its bottom part, since the only remaining 4-cycles are confined to $\demx$ and $\demz$.}
    \label{tab:cycles}
    \resizebox{\linewidth}{!}{
    \begin{tabular}{|c|c|c|c|c|c|c|c|c|c|c|c|c|c|}
    \hline
     \multicolumn{1}{|c|}{\multirow{2}{*}{Code}} & \multicolumn{3}{c|}{$\demx$} & \multicolumn{3}{c|}{$\demz$} & \multicolumn{3}{c|}{$\demxyz$} & \multicolumn{3}{c|}{Bottom part of $\demxyzbar$} \\
    \cline{2-13}& Size & $\bar{w}_r$ & 4-cycles &  Size & $\bar{w}_r$ & 4-cycles & Size & $\bar{w}_r$ & 4-cycles & Size &  $\bar{w}_r$ & 4-cycles\\
    \hline
    $[[72,12,6]]$     & 180 $\times$ 1800 & 33.2 & 10440   & 252 $\times$ 2232  & 30.86 & 13248 & 432 $\times$ 16164  & 210.92 & 2628756  & 4032 $\times$ 20196  & 8.02 & 0\\
    $[[90,8,10]]$     & 405 $\times$ 4050 & 34.0 & 24030   & 495 $\times$ 4590  & 32.36 & 27720 & 900 $\times$ 34965  & 223.35 & 5967945  & 8640 $\times$ 43605  & 8.09 & 0\\
    $[[144,12,12]]$   & 792 $\times$ 7920 & 34.18 & 47232   & 936 $\times$ 8784  & 32.77 & 53280 & 1728 $\times$ 67752 & 226.46 & 11584296 & 16704 $\times$ 84456 & 8.11 & 0\\
    \hline
    \end{tabular}
    }
\end{table*}

\subsubheading{The Decoding Graph Perspective} 
\arxiv{\label{sec:graph}}
Let us first consider an arbitrary decoding graph that contains 4-cycles (the connection with the previous section will be made later), which we aim to eliminate to enable more effective MP inference.  This necessarily involves modifying the decoding graph, while ensuring that the resulting graph corresponds to a decoding problem equivalent to the original one. Moreover, since 4-cycles can combine to form \emph{bicliques} -- i.e., complete bipartite subgraphs of the decoding graph -- these are precisely the types of structures we aim to eliminate. A 4-cycle is illustrated in Fig.~\ref{fig:4-cycle}, and a biclique is shown in Fig.~\ref{fig:biclique}. Any two error nodes and any two check nodes participating in the biclique determine a 4-cycle. Note that a biclique is necessarily an induced subgraph, formed from one subset of error nodes and one subset of check nodes, along with all edges from the decoding graph connecting pairs of nodes in these subsets. In terms of the decoding matrix, a biclique corresponds to a submatrix with all entries equal to $1$. 

The GARI procedure is illustrated in Fig.~\ref{fig:gari}, and it is applied to maximal bicliques in the decoding graph -- i.e., bicliques that cannot be extended by adding any additional error or check nodes. We modify the decoding graph by adding a new error node (in blue), representing the sum of all error nodes in the biclique. We enforce this last condition by also introducing a new check node (in blue) connected to the error nodes in the biclique and the newly added one, with its syndrome value set to zero. Finally, for each check node in the biclique, we replace its corresponding incident edges with a single edge connecting it to the newly added error node. We note that the above procedure was previously developed, but applied only to 4-cycles, in \cite{yedidia2002generating, fossorier2015polar} for decoding some families of classical linear block codes under belief propagation.

Clearly, the above procedure eliminates all 4-cycles forming the biclique. While new 4-cycles may appear involving the  nodes introduced by GARI, it can be easily seen that their number is reduced. Thus, recursively applying the procedure will eventually eliminate all 4-cycles in the graph.

We also note that, although the above procedure augments the decoding graph with additional error and check nodes, it increases the number of edges by only one when applied to a 4-cycle, preserves the number of edges when applied to a 5-node biclique, and strictly reduces the number of edges in all other cases. This can be easily seen by comparing the number of edges in Fig.~\ref{fig:biclique} and Fig.~\ref{fig:gari}. Consequently, the augmented graph may contain fewer edges than the original graph, implying that an MP decoder operating on the augmented graph may incur a reduced computational load (i.e., performs fewer message updates per iteration) compared to the same decoder operating on the original graph. This is typically the case for correlated detector error model matrices $\demxyz$, due to the generally large size of the maximal bicliques (see also the BB code examples below).

We now claim that, for the correlated detector error model discussed in the previous section, the GARI matrix $\demxyzbar$ is the adjacency matrix of the graph 
produced by one possible application of the GARI transform to the decoding graph with adjacency matrix $\demxyz$. 
To see this, we first note that any column of $\demx$ or $\demz$ that is repeated one or more times in $\demx'$ or $\demz'$, respectively, determines a biclique in $\demxyz$. For instance, assuming that the $i$th column of $\demx$  equals the columns $j_1$ and $j_2$ in $\demx'$, we obtain a biclique with three error nodes as in Fig.~\ref{fig:biclique}: the three error nodes correspond to these columns, while the check nodes (left unlabeled in the figure) correspond to the rows containing 1 entries in them. Upon application of GARI, the newly introduced error and check nodes (depicted in blue in Fig.~\ref{fig:gari}) correspond, respectively, to the $i$th column within the $\bar{e}_Z$ column block and the $i$th row within the third row block of $\demxyzbar$. The edges connecting the newly added check to the error nodes $j_1$ and $j_2$ correspond to 1 entries in the $U$ matrix, while the edges connecting it to the white and blue error nodes labeled $i$ correspond to the 1 entries in the two identity matrices on the left and right sides of $U$. The edges connecting the newly added error node to the four check nodes from the original graph reproduce the corresponding $D_X$ column.

Summarizing, upon applying GARI, 
the added error nodes  correspond to columns $\bar{e}_Z$ and $\bar{e}_X$ in~\eqref{eq:gari-matrix}, while the added check nodes correspond to the bottom part of $\demxyzbar$, containing the $I$'s, $U$ and $V$ submatrices. $D_X$ and $D_Z$ submatrices in the top part of $\demxyz$ are reproduced by the new edges added between the original check nodes and the newly introduced error nodes, and they contain the only remaining 4-cycles in the new graph. See Table~\ref{tab:cycles} for a quantitative summary of the size, average row weight, and number of 4-cycles in $\demx$, $\demz$, $\demxyz$, and $\demxyzbar$ for various BB codes.  The number of non-zero entries in each matrix can be computed as the product of the number of rows and the average row weight. For instance, for the [[144, 12, 12]] code, the $\demxyz$ matrix has $1728\times 226.46 = 391323$ non-zero entries, while the $\demxyzbar$ matrix (including $\demx$, $\demz$, and the bottom part) has  $792\times 34.18 + 936\times 32.77 + 16704\times 8.11 = 193213$ non-zero entries, which is less than half the number of non-zero entries in $\demxyz$.

As discussed above, the GARI transform can be applied recursively to further eliminate the remaining 4-cycles. However, for the tested BB codes\arxiv{ (see Section~\ref{sec:accuracy})}, the achieved performance was on par with or exceeded that of significantly more complex state-of-the-art decoders, and at the time of writing this paper, we did not pursue further optimization.

We also note that removing only the cycles through $Y$-type error nodes enables a more accurate treatment of correlations, without optimizing the decoding of a single error type. Indeed, a closer examination of the GARI decoding matrix in~\eqref{eq:gari-matrix},  reveals that a GARI MP decoder can be interpreted as two decoders operating on the single-detector-type matrices $\demx$ and $\demz$, which exchange information through the bottom part of  $\demxyzbar$. This information exchange allows the decoder to account for correlations between $X$, $Y$, and $Z$ errors. Such a decoder architecture is further discussed and illustrated in~\natcomm{a subsequent section}\arxiv{Section~\ref{sec:hw-architecture}}.

\subheading{Randomized Scheduling based Ensemble Decoding}

Now that we have a graph enabling improved MP inference, we can apply MP decoding on it. It is to be expected that any MP decoder will perform better on the GARI graph than on the original one. However, optimizing accuracy and speed requires a judicious choice of the MP decoder, encompassing both the message update rules and the message update schedule. 
We refer to the \natcomm{Methods section}\arxiv{Appendix} for a discussion on different message-update schedules (flooding, serial, layered) and, in the following, discuss the choice of MP decoding for GARI and the ensemble decoding approach.

\subsubheading{MP Decoding on GARI}
\arxiv{\label{sec:mp-decoding-on-gari}}
We consider the normalized variant of the MS decoder (NMS), in which a scaling (normalization) factor is applied to check node messages to improve the accuracy of inference on loopy graphs~\cite{chen2005reduced}. 
It is also the preferred choice for hardware implementations, owing to its simple arithmetic and robustness when messages are quantized with a small number of bits. Lowering the bit-width of the exchanged messages is important both for reducing footprint and for increasing speed~\cite{nguyen2017analysis}.  

We also employ a horizontal hybrid serial-layered schedule, as explained in the next paragraph. Horizontal schedules are often the \emph{de facto} choice in hardware implementations due to their efficiency in resource allocation and straightforward integration into processing pipelines~\cite{boutillon2014hardware}. We note, however, that we also experimented with various vertical schedules, which proved less effective than horizontal ones in terms of decoding accuracy and speed. In the following, all schedules are  horizontal. 

\emph{Hybrid schedule:} We  consider the GARI decoding matrix $\demxyzbar$ given in Eq.~\eqref{eq:gari-matrix}. For the top part of $\demxyzbar$, consisting of the rows corresponding to the $\demx$ and $\demz$ blocks, we use a randomized serial schedule.  For the bottom part, consisting of the remaining rows, we use a layered schedule with two layers: the first layer contains the rows corresponding to the $U$ matrix, and the second contains those corresponding to the $V$ matrix. 

For the top part, since the only non-zero entries lie within the  $\demx$ and $\demz$ matrices, which do not share any rows or columns, we can equivalently use two MP decoders running in parallel -- one for $\demx$ and one for $\demz$ -- each employing a randomized serial schedule. At each iteration, these decoders exchange information via the processing performed in the bottom part of the $\demxyzbar$.

For the bottom part, the fact that the layers are well defined (i.e., rows within each layer do not intersect) follows from the fact that $U$ and $V$ have column weight~$1$\arxiv{, as discussed in Section~\ref{sec:gari}}. Since there are only two layers and processing the same layer twice in succession (without processing the other layer in between) would not change the outgoing messages, randomization is not possible. Thus, we simply process the layers in their natural order. Note that a serial schedule processing the rows of the bottom matrix in their natural order would perform exactly the same computations (i.e., produce the same messages). Thus, layered scheduling is used in the bottom matrix not to improve accuracy, but merely to reduce latency.  

\emph{Prior probabilities:} The MP decoder is initialized with a prior probability value for each column of $\demxyzbar$ (i.e., each error node, in graph terminology). For the $e_z$, $e_x$, and $e_Y$ columns in~\eqref{eq:gari-matrix}, we use the prior probabilities obtained from the detector error model (as done for $\demxyz$ in~\eqref{eq:demxyz-matrix}). For columns $\bar{e}_Z$ and $\bar{e}_X$, corresponding to new error nodes added by GARI, no prior knowledge is available, so their  prior probabilities are set to $0.5$ (corresponding to a log-likelihood ratio of $0$). One might object that prior probabilities could be obtained from Eq.~\eqref{eq:bar-errors}. However, this is precisely what the MP decoder does after the first decoding iteration. Manually setting prior probabilities is not only unnecessary but may be detrimental, as it  duplicates the information obtained through MP inference. Finally, since $\bar{e}_Z$ and $\bar{e}_X$ columns have no prior knowledge, we start the decoder by processing the bottom part of $\demxyzbar$ first, followed by the top part (starting with the top would simply waste the first iteration on that part).

\emph{Early stopping:} 
Let us consider a memory experiment, in which we protect logical-$Z$ observables (i.e., the protected logical state is a tensor product of eigenstates of the logical-$Z$ operators). In the detector error model, the physical qubits are measured in the $Z$ basis at the end, which allows us to detect logical-$X$ errors, manifested as changes in the protected logical observables. Although our decoder exploits correlations and effectively estimates the error vector $(e_Z, e_X, e_Y, \bar{e}_Z, \bar{e}_X)$, we can actually only assess the accuracy of the $X$ error estimation,  including contributions from $Y$ errors -- that is $\bar{e}_X$.  Thus, for early stopping of our MP decoder, we only check whether the $Z$-type syndrome is satisfied, that is, $D_Z \bar{e}_X = s_Z$. This is checked after each decoding iteration, and the MP decoder stops when the $Z$-type syndrome is satisfied or when a maximum number of iterations is reached. Similarly, for the $X$-memory experiment, one would implement the early stopping on $D_X$ instead.

As an aside, we note that we also experimented with a different early stopping criterion, checking both $Z$ and $X$ syndromes, that is $D_Z \bar{e}_X = s_Z$ and $D_X \bar{e}_Z = s_X$ (while, of course, we could estimate only the logical-$X$ error rate, based on the decoded $\bar{e}_X$). This yielded a similar logical-$X$ error rate but increased the average number of decoding iterations by approximately a factor of two, since the decoder needs to converge on both  $D_Z$ and $D_X$ before it stops. Also, convergence on $D_Z$ generally occurs faster, aided by the final measurements of physical qubits in $Z$ basis. Importantly, such a stopping criterion could be used in experiments in which both $X$ and $Z$ logical operators must be preserved. However, our focus is on memory experiments involving the preservation of a single type of logical operator (in line with current practice in the literature), and we aim to show how its preservation can be enhanced by effectively exploiting correlations among $X$, $Y$, and $Z$ errors.

\subsubheading{Ensemble Decoding Approach}
\arxiv{\label{sec:ensemble_decoding}}
The benefits of randomization in scheduling can be further leveraged through an ensemble decoding approach. This method involves running multiple MP decoder instances in parallel, each decoding the GARI matrix but using a distinct randomization seed for the serial part of the hybrid serial-layered schedule.

\emph{Early ensemble stopping:}  Since our decoders run in parallel and may converge after different numbers of iterations, we also need a global strategy to stop the entire ensemble. We adopt a latency-minimizing strategy, in which all MP decoder instances stop as soon as one of them converges.  If several MP decoders converge simultaneously (i.e., after the same number of decoding iterations), we select the most likely error among those estimated by these decoders, where the likelihood of an error is determined according to the prior probabilities.
However, recall that the MP decoder stops when the $Z$-type syndrome is satisfied, $D_Z \bar{e}_X = s_Z$, and that no prior probabilities for $\bar{e}_X$ has been provided to the decoder. Fortunately, as mentioned earlier, we can use Eq.~\eqref{eq:bar-errors} to compute prior probabilities for $\bar{e}_X$ (which are identical to those obtained from the $Z$-type only detector error model).  While these probabilities are not needed by the decoder (see earlier discussion), we use them to compute the likelihood of each estimated error and then select the most likely one.

\subheading{Hardware Architecture}
\arxiv{\label{sec:hw-architecture}}

\begin{figure}[!t]
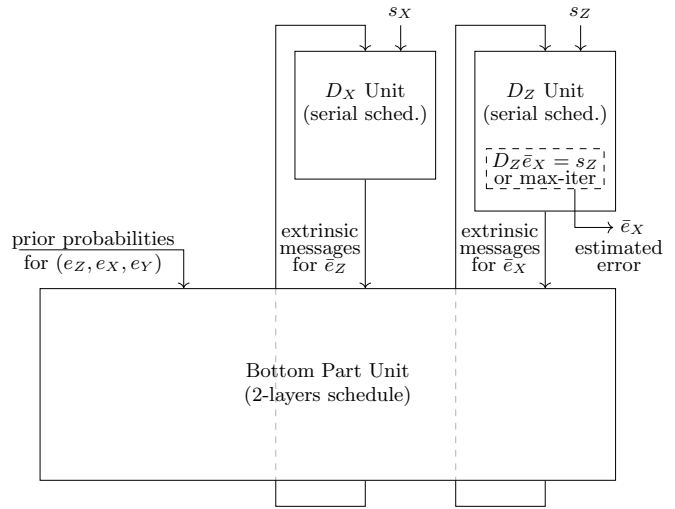

\hspace*{-2ex}
\includestandalone[width=\linewidth+2ex]{hw_archi_v3} 
\caption{High-level architecture of the GARI MP decoder. MP processing moves upward, completing one iteration after first updating the bottom-part unit, followed by the $D_X$ and $D_Z$ units, which are updated in parallel. The input consists of prior probabilities for $(e_Z, e_X, e_Y)$, and the output of the estimated $\bar{e}_X$ error.}
\label{fig:hw-archi}
\end{figure}

The GARI MP decoder features a hardware-efficient architecture, which we discuss in this section. The architecture is illustrated in Fig. \ref{fig:hw-archi}, and consists of three distinct processing units.
The $D_X$ and $D_Z$ units implement MP decoding on the $D_X$ and $D_Z$ submatrices, respectively, and they exchange information through a third processing unit that implements MP decoding on the bottom part of $\demxyzbar$. We note that this high-level architecture is generic and can accommodate different MP decoders, including different message update rules and scheduling strategies. 
Below, we examine in detail the benefits of the hybrid serial–layered schedule discussed in the previous section.

\emph{Resource efficiency :} Serial schedules are implemented in hardware using a single check node processor, which is reused to sequentially process all the checks (rows) of the decoding matrix. Thus, implementing  $\demx$ and $\demz$ units requires a single check-node processor for each. Considering an ensemble decoding approach, the number of check-node processors per $\demx$ or $\demz$ unit equals the size of the ensemble. Even in this case, the number of check-node processors remains significantly lower than that required for the parallel implementation of a single MP decoder on either $\demx$ or $\demz$, which equals the number of rows of the corresponding matrix. This is particularly important as the code distance and thus the size of the decoding matrix increase  (see, for instance, Table~\ref{tab:cycles}). 
While such an increase does not affect the resource usage of $D_X$ and $D_Z$ units in our architecture,  it can hinder the implementation of a fully parallel schedule due to the limited resources available on FPGAs.

The most resource-intensive part of our architecture is the implementation of the bottom-part unit, which requires a number of check-node processors equal to the maximum layer size (i.e., the maximum number of rows between the first and second layers). However, the weights of the rows in the bottom part of $\demxyzbar$ are considerably smaller than those of the $\demx$ and $\demz$ rows (see Table~\ref{tab:cycles}). Thus, check node processors in the bottom-part unit require fewer resources than in the $D_X$ and $D_Z$ units, so increasing their number is less of a concern.

\emph{Low latency:} The serial architecture of $D_X$ and $D_Z$ units reduces routing complexity, since a single check node processor is connected to a single bank of registers and memory. 
Such an architecture can operate at high clock frequencies, thereby compensating for the inherent latency induced by the sequential nature of the serial schedule.  Moreover, the horizontal serial schedule considered in our architecture provides a significant speedup compared to the vertical schedule, due to the ratio between the number of rows and columns in the $D_X$ and $D_Z$ matrices (at least ninefold speedup for the BB codes, see Table~\ref{tab:cycles}). By way of comparison, it is worth noting that a parallel implementation would require multiple wiring connections (proportional to the edges of the graph), which could cause routing congestion and contribute significantly to the critical path delay in FPGAs that have fixed routing lines. 

The layered schedule in the bottom part unit may reduce the operating clock frequency, due to the increased footprint and routing complexity. However, this effect is mitigated by the reduced logic depth of the check-node processors, while the inherent parallelism of the schedule enables a significant latency reduction. 

Finally, we note that processing units can employ pipelining to reduce the critical path and increase the maximum achievable clock frequency\arxiv{, as discussed in Section~\ref{sec:decoding-latency}}. The pipelined implementation also supports hardware sharing, allowing resources to be reused across different decoding runs, thereby further improving resource efficiency. Resource sharing across decoding runs was not implemented in the present work, but could be explored in future research.

\arxiv{\heading{Results}}
\natcomm{\subheading{Numerical Results on Decoding Accuracy}}

We consider memory experiments, in which all logical qubits are initialized in an eigenstate of their corresponding logical-$Z$ operators. To benchmark our method, we consider the BB codes with parameters [[72,12,6]], [[90,8,10]] and [[144,12,12]] from~\cite{bravyi2024high}. 
For a code of distance $d$, we consider the circuit obtained by repeating the syndrome measurement $d$ times, followed by a final measurement of the data qubits in $Z$ basis. The corresponding detector error model is constructed and simulated for a given noise model using Stim~\cite{Gidney_2021}. Note that detector circuits may not be distance preserving (i.e., the circuit level distance may be less than $d$), but we will systematically refer to them using the $[[n,k,d]]$ notation of the corresponding code. 

\arxiv{\subheading{Decoding Accuracy}\label{sec:accuracy}}

We consider various decoders operating on different decoding matrices, namely $\demz$, $\demxyz$, and $\demxyzbar$ (i.e., the GARI matrix). The goal is to provide a comparison of our method with BPOSD decoder, as well as with more recent, higher-accuracy decoders such as Relay-BP and Tesseract. We discuss below the hyperparameters used in our simulations, where decoder names are prefixed with the decoding matrix they operate on (using GARI for $\demxyzbar$). 

\emph{GARI-NMS:} NMS decoder with hybrid serial-layered schedule, operating on the $\demxyzbar$ matrix\arxiv{, as detailed in Section~\ref{sec:mp-decoding-on-gari}}. The serial schedule is randomized and the maximum number of iterations is set to 400. 
The NMS normalization factor is set to either $\sum_{i=1}^5 1/2^i = 0.96875$ or $\sum_{i=1}^7 1/2^i = 0.9921875$, chosen to optimize accuracy while facilitating hardware implementation using bit shifts and adders. 
The GARI-NMS-ensemble decoders employ either 24 or 48 GARI-NMS decoders in parallel for the uniform depolarizing noise and  SI1000 noise models, respectively (discussed below). We denote these decoders as GARI-NMS-$\times s$, where $s$ is the size of the ensemble. Ensemble decoding is only used for BB codes with distances 10 and 12.

\emph{($\demz$/$\demxyz$)-NMS:}  For comparison purposes, we also consider an NMS decoder with a randomized serial schedule, using a maximum number of 400 decoding iterations and normalization factor of 0.96875, but operating on either $\demz$ or $\demxyz$ matrix.

\emph{($\demz$/$\demxyz$)-BPOSD:} 
We use the open-source library ~\cite{Roffe_LDPC_Python_tools_2022,roffe2020decoding} to simulate the order-zero BPOSD decoder. We fine-tune BPOSD decoders on both $\demz$ and $\demxyz$ using different NMS normalization factors, schedules and number of iterations, and then use the best resulting decoder for comparison with our method \natcomm{(see Methods for details)}\arxiv{(see Appendix\ref{app:BPOSD_sr_vs_pl} for details)}. 
We obtain the best performance when combining OSD post-processing with an NMS decoder operating on $\demxyz$ for distance $d=6$, but on $\demz$ for distances $d=10$ and $d=12$, using a normalization factor of 0.625, randomized serial scheduling, and a maximum number of 30 iterations (a larger number of iterations degrades performance at low physical error rates due to erratic behavior of the MP decoder). We note that the NMS decoder in BPOSD library~\cite{Roffe_LDPC_Python_tools_2022} uses a vertical serial schedule  (in contrast to horizontal schedules used in our implementation of the NMS).

\begin{figure*}
\centering
    \begin{subfigure}[b]{0.45\textwidth}
    \captionsetup{format=centercaption}
    \setlength{\abovecaptionskip}{0pt}
        \centering
        \includegraphics[width=\textwidth]{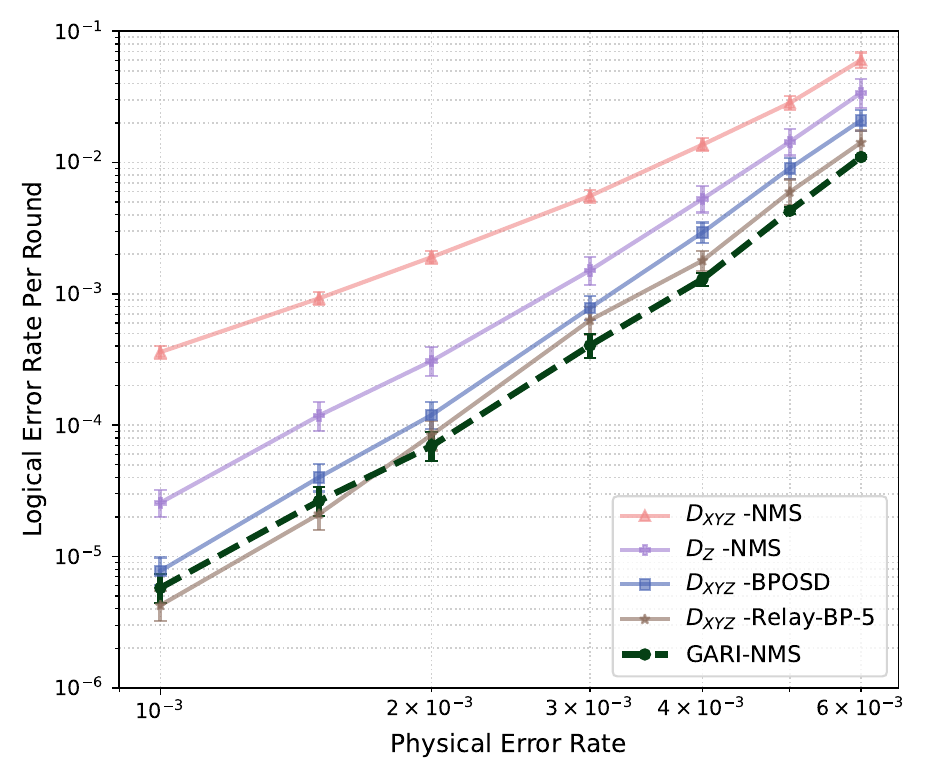}
        \caption{[[72,12,6]] code, Stim circuit from~\cite{gong2024lowlatencyiterativedecodingqldpc}}
        \label{fig:accuracy-d6}
    \end{subfigure}
    \hfill
    \begin{subfigure}[b]{0.45\textwidth}
    \captionsetup{format=centercaption}
    \setlength{\abovecaptionskip}{0pt}
        \centering
        \includegraphics[width=\textwidth]{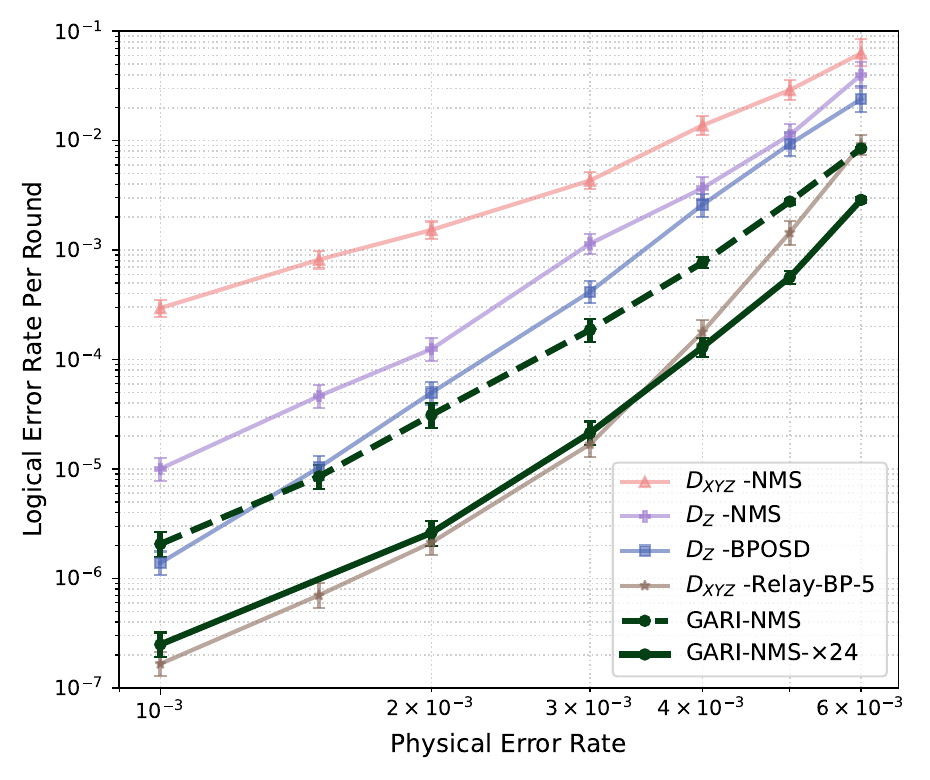}
        \caption{[[90,8,10]] code, Stim circuit from~\cite{gong2024lowlatencyiterativedecodingqldpc}}
        \label{fig:accuracy-d10}
    \end{subfigure}

    \vskip 3mm  

    \begin{subfigure}[b]{0.45\textwidth}
    \captionsetup{format=centercaption}
    \setlength{\abovecaptionskip}{0pt}
        \centering
        \includegraphics[width=\textwidth]{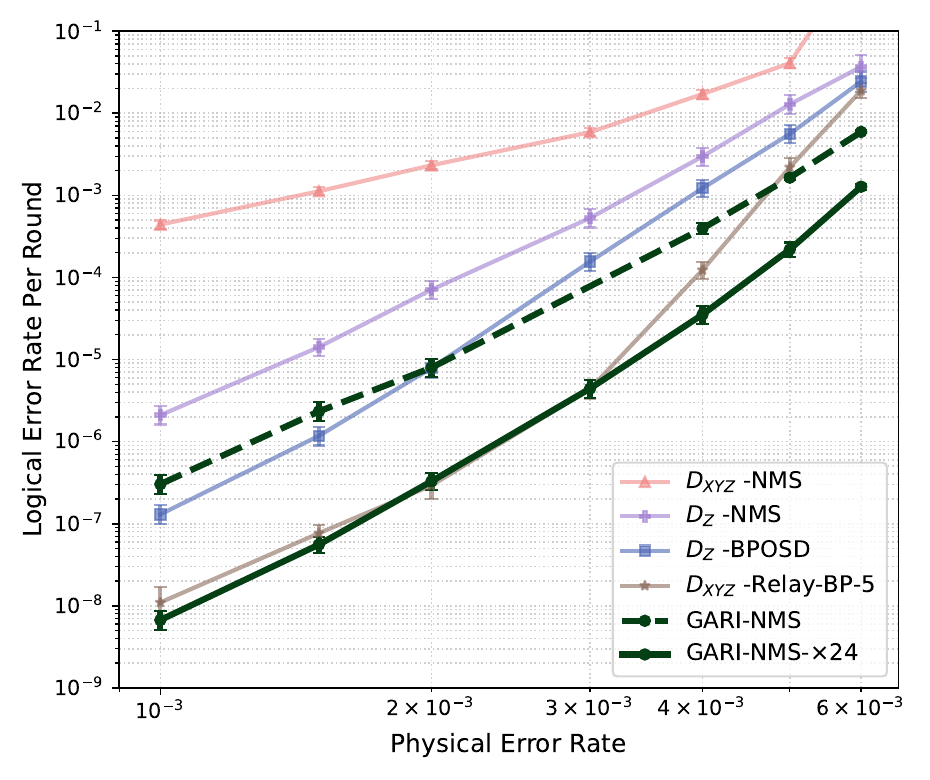}
        \caption{[[144,12,12]] code, Stim circuit from~\cite{gong2024lowlatencyiterativedecodingqldpc}}
        \label{fig:accuracy-d12}
    \end{subfigure}
    \hfill
    \begin{subfigure}[b]{0.45\textwidth}
    \captionsetup{format=centercaption}
    \setlength{\abovecaptionskip}{0pt}
        \centering
        \includegraphics[width=\textwidth]{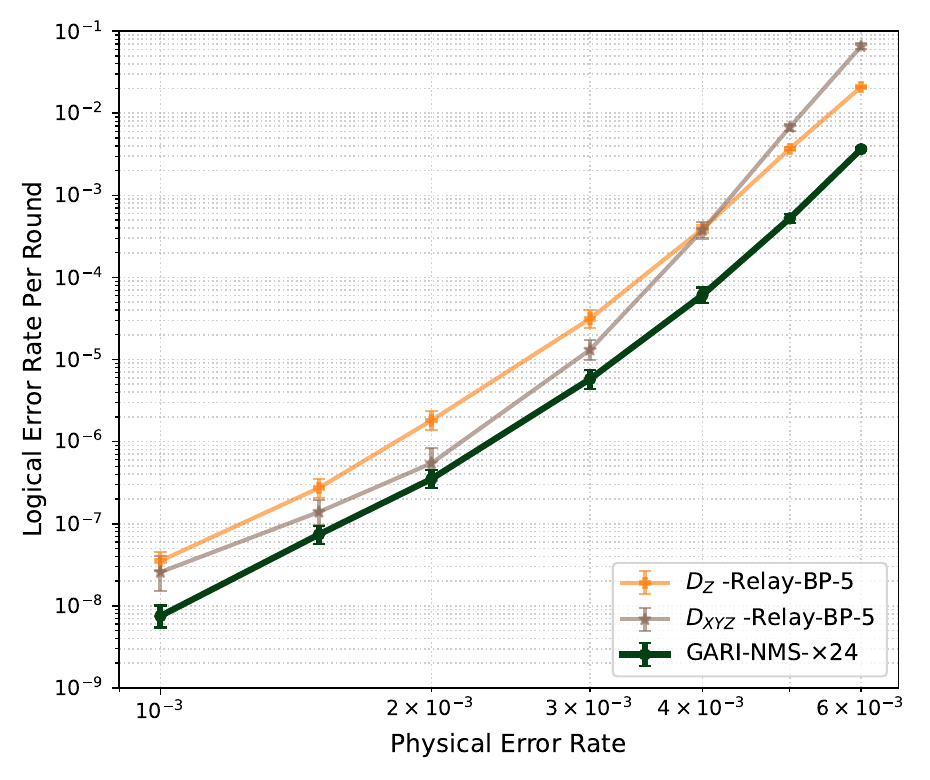}
        \caption{[[144,12,12]] code, Stim circuit from~\cite{github_relaybp}}
        \label{fig:accuracy-d12-ibm}
    \end{subfigure}
    \caption{Logical error rate per-round vs. physical error rate for various BB codes under uniform depolarising circuit-level noise, using Stim circuits from either~\cite{gong2024lowlatencyiterativedecodingqldpc} or~\cite{github_relaybp}. We prefix decoder names with the decoding matrix they operate on, using GARI for $\protect\demxyzbar$. The hyperparameters of the decoders are discussed in the main text and error bars indicate 99\% confidence intervals. Note that we consider the best performing BPOSD decoder, operating on either $\demxyz$ or $\demz$, depending on the code, as explained in the main text. For GARI-NMS decoders, the normalization factor is set to 0.96875 in subfigures~(a)-(c), and 0.9921875 in~(d). For codes with distances 10 and 12, the GARI-NMS-ensemble decoder uses 24 GARI-NMS decoders in parallel.}
    \label{fig:accuracy}
\end{figure*}

\emph{$\demxyz$-Relay-BP:} 
We use the open-source library~\cite{github_relaybp}, associated with the paper~\cite{muller2025improvedbeliefpropagationsufficient}, to simulate Relay-BP, using the hyperparameters specified in the library. We compare our results with the XYZ-Relay-BP-5 (operating on $\demxyz$) decoder, which handles correlated errors and has been shown to outperform BPOSD. It uses  301 DMem-BP decoders in series with up to 18080 decoding iterations. The decoder searches for five different solutions (thus the suffix 5 in the notation) and returns the one with the lowest weight. We note that the simulation results in~\cite{muller2025improvedbeliefpropagationsufficient} use 601 DMem-BP decoders in series, instead of 301, which is the default value in the library. We noticed this rather late and ran several simulations with 601 DMem-BP decoders, but they were too slow to produce meaningful results, especially at low physical error rates.

\emph{Logical error rates:} We determine the logical error rate as the ratio of decoding failures (i.e., either the decoder did not converge, or at least one logical observable is flipped by the decoded error) to the total number of decoded runs, i.e., LER = decoding failures/number of samples. The logical error rate per syndrome measurement round (in short, per-round) is obtained by~\cite{beni2025tesseractsearchbaseddecoderquantum},
\begin{equation}
\label{eq:ler_per_round}
     \text{LER}_\text{round} = (1-(1-2 \times \text{LER})^{1/r})/2.
\end{equation}
Unless otherwise stated, all reported logical error rates are per-round. Error bars in all plots indicate 99\% confidence intervals, calculated from at least 100 decoding failures, except for the Relay-BP and Tesseract (discussed later) decoders, for which the number of decoding failures is about 40 at low physical error rates.

\medskip \emph{Uniform depolarizing noise:} Fig.~\ref{fig:accuracy} presents the per-round logical error rate results under the circuit-level uniform depolarizing noise model. The detector error model is instantiated using the Stim circuits from~\cite{gong2024lowlatencyiterativedecodingqldpc} for Figs.~\ref{fig:accuracy-d6}–\ref{fig:accuracy-d12}, and the Stim circuit from the Relay-BP library~\cite{github_relaybp} for Fig.~\ref{fig:accuracy-d12-ibm}.

 First, we observe that the $\demxyz$-NMS decoder performs significantly worse than $\demz$-NMS. This degradation is due to the large number of 4-cycles in $\demxyz$ introduced by $Y$-type errors, which hinder the NMS decoder's ability to handle correlations between Pauli errors. Table~\ref{tab:cycles} lists the number of 4-cycles in these matrices, as well as the average check degree, for all the codes considered. It is worth noticing that the bottom part of the GARI matrix contains no 4-cycles and has a smaller average check degree compared to $\demxyz$,  making it more suitable for MP decoding. As a result, the GARI-NMS  decoder not only outperforms  $\demz$-NMS, but  it is already on par with BPOSD for BB codes of distance $d=10$ and $d=12$, while being significantly more lightweight and faster.  Moreover, for the smaller distance $d=6$ code, GARI-NMS outperforms BPOSD, achieving performance similar to that of the $\demxyz$-Relay-BP-5 decoder.

We now consider the GARI-NMS-$\times 24$ ensemble decoder, and report its decoding performance for the BB codes of distance $d=10$ and $d=12$. Let us note that for the BB code of distance $d=6$, the ensemble decoder does not further improve performance.
For distances $d=10$ and $d=12$, we can observe that GARI-NMS-$\times 24$ and $\demxyz$-Relay-BP-5 decoders achieve similar performance, with the former performing slightly better at high physical error rates or when using the Stim circuit from the Relay-BP library, as shown in Fig.~\ref{fig:accuracy}(d). 

While we prioritized minimizing latency, the GARI-NMS-ensemble performance could potentially be improved by waiting for multiple decoders to converge, and then selecting the most likely (e.g., lowest weight) solution. This echoes the idea of Relay-BP-5, which stops after five decoders have converged. However,  while both approaches rely on ensemble decoding, it is worth noting that they differ in a key aspect. Our method runs all decoders in the ensemble in parallel, with each operating independently. In contrast, Relay-BP executes the decoders sequentially, relaying information from one decoder to the next. This serial operation increases decoding latency and limits the ability to exploit the parallel processing capabilities of CPUs and FPGAs.

Finally, for completeness of the comparison, we also discuss the case of the $\demz$-NMS-ensemble decoder (not shown in Fig.~\ref{fig:accuracy} to maintain clarity), where the randomized-scheduled ensemble decoding approach is applied to $\demz$-NMS rather than the GARI-NMS decoder. For distance $d=12$ and physical error rate $10^{-3}$, the $\demz$-NMS-$\times 24$ decoder achieves a logical error rate of $1.6\times10^{-7}$, on par with BPOSD. This highlights that it is the combination of GARI and randomized-scheduled ensemble decoding that achieves the best accuracy.

\medskip\emph{SI1000 noise model and Tesseract:}
For comparison with Tesseract, we use the data provided by the authors of~\cite{beni2025tesseractsearchbaseddecoderquantum}. The simulations in that work employ the SI1000 noise model and a different gate set (CZ-H-MZ) for BB code circuits. Accordingly, we simulate our decoder under the same noise conditions and on the same circuits, which are available in the open-source library\arxiv{~\cite{github_tesseract}} 
associated with~\cite{beni2025tesseractsearchbaseddecoderquantum}.

\begin{figure}[!t]
\includegraphics[width=.45\textwidth]{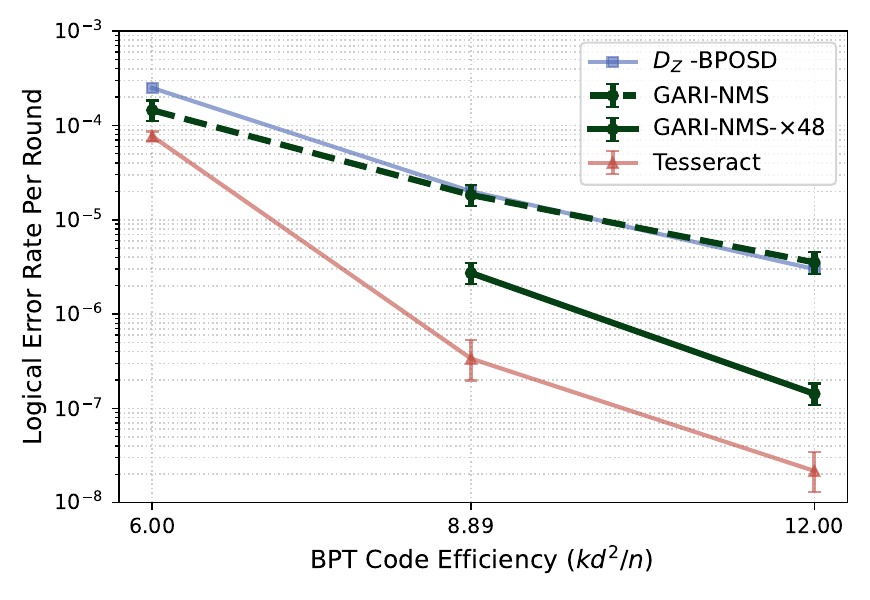}
\caption{Logical error rate performance of the proposed method compared with the Tesseract decoder introduced in ~\cite{beni2025tesseractsearchbaseddecoderquantum}, for the SI1000 noise model with a physical error rate of 0.001. 
We plot the logical error rate versus the BPT code efficiency $k d^2 / n$, for BB codes with parameters $[[72,12,6]]$, $[[90,8,10]]$, and $[[144,12,12]]$, corresponding to efficiencies of 6, 8.89, and 12, respectively (see also Fig.~\ref{fig:logical-error-rate-comp}). Error bars indicate 99\% confidence intervals. The data for the BPOSD and Tesseract decoders are taken from~\cite{beni2025tesseractsearchbaseddecoderquantum}. GARI-NMS uses a normalization factor of 0.9921875, and ensemble decoders for distances 10 and 12 use 48 GARI-MP decoders in parallel. }
\label{fig:logical-error-rate-google_si1000}
\end{figure}

In this setting, the GARI-NMS-ensemble decoder achieves an accuracy within an order of magnitude of Tesseract, as shown in Fig.~\ref{fig:logical-error-rate-google_si1000}. However, Tesseract is a graph search–based decoder that searches through a graph of all error subsets to find the lowest-cost subset consistent with the input syndrome. Although the search can be made more efficient using the A* algorithm and pruning heuristics (since the search graph is exponentially large), its complexity remains considerably higher than that of MP decoders. To the best of our knowledge, no other MP-based decoding solution has been reported to achieve a comparable level of decoding performance in this setting.

\arxiv{\subheading{Real-Time Decoding}\label{sec:decoding-latency}}
\natcomm{\subheading{Real-Time Decoding Implementation}}

We report implementation results obtained for the Virtex UltraScale+ VU19P (16\,nm), currently the highest-capacity FPGA in production by AMD.
The code under test is the [[144,12,12]] BB code. The cores for the $D_X$, $D_Z$, and bottom-part units were implemented independently using Vivado 2021.2. 

The serial $D_X$ and $D_Z$ processing units employ 10 pipeline stages and operate at a maximum clock frequency of $357$\,MHz. This corresponds to a clock period of 2.8\,ns, with 60\% of the delay due to routing and 40\%  to logic gates.
Consequently, the per-iteration delay is $2.22\,\mu\text{s}$ for $D_X$ and $2.62\,\mu\text{s}$ for $D_Z$. 
 For the bottom part processing unit, the maximum clock frequency is over 20 times lower than that of the serial units, due to the high degree of parallelism and the absence of pipelining. 
The total delay 
 for processing both layers is $272$\,ns, with 80\% of the delay due to the routing. 
 The total delay per iteration for the GARI MP architecture is the sum of the maximum delay of $D_X$ and $D_Z$ units (which run concurrently), and the delay of the layered bottom part unit. This results in a total per-iteration delay of $2.9\,\mu\text{s}$.  We note that, in practice, a fully integrated design with proper clock-domain crossing is expected to achieve similar decoding latency, with only negligible overhead due to synchronization. In addition, if multi-FPGA is required, high-speed transceivers such as GTY included in V19P, with 4.5 Tb/s of bandwidth and tens of nanoseconds of latency in raw mode can be used to preserve the timing behaviour of the independently implemented modules.

\begin{table}[!t]    
    \centering
    \caption{Average number of decoding iterations performed by GARI-NMS  ($s=1$) and GARI-NMS-ensemble  (ensemble size $s=24$) decoders.}
    \label{tab:iterations}
    \begin{tabular}{|c|c|c|c|c|c|}
    \hline
    \multicolumn{1}{|c|}{\multirow{2}{*}{PER}} 
     & [[72,12,6]] & \multicolumn{2}{c|}{[[90,8,10]]} & \multicolumn{2}{c|}{[[144,12,12]]} \\
    \cline{2-6} & $s=1$  & $s=1$ & $s=24$ & $s=1$ & $s=24$ \\
    \hline
    0.001 & 1.36    &  1.75   &  1.04  &  2.28   &  1.13   \\
    0.002 & 2.26    &  3.32   &  1.56  &  4.13   &  2.2    \\
    0.003 & 3.72     &  5.93   &  2.71  &  6.51   &  3.53   \\
    0.004 & 6.64    &  11.5   &  4.61  &  11.97  &  5.31   \\
    0.005 & 13.89   &  25.33  &  8.94  &  25.25  &  9.18    \\
    0.006 & 29.31  &  55.37  &  22.49 &  57.38  &  20.79   \\
    \hline
    \end{tabular}
\end{table}

Assuming a time budget of $1\,\mu\text{s}$ per syndrome measurement round~\cite{Battistel_2023}, and given that decoding is performed across a number of rounds equal to the code distance,  the resulting time budget for decoding the BB code under test is $12\,\mu\text{s}$. Hence, on average, slightly more than four decoding iterations can be accommodated within the given time budget. According to the average number of decoding iterations reported in Table~\ref{tab:iterations},  the  GARI MP ensemble decoder operates in real-time (i.e., within the specified time budget) for physical error rates below 0.003. 

\begin{figure}[!t]
    \centering
    \includegraphics[width=.48\textwidth]{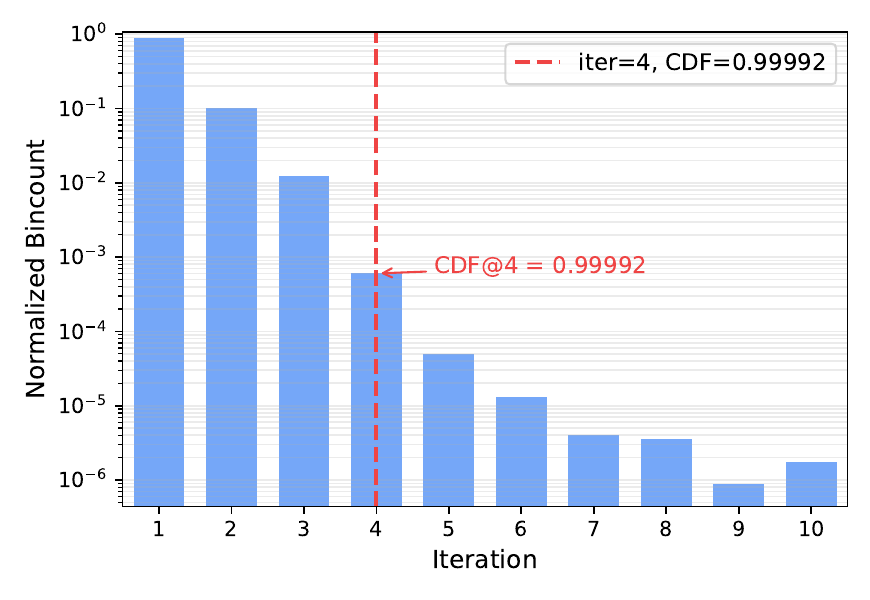}
    \caption{Percentage of decoding runs ($y$-axis) converging at a given iteration number ($x$-axis) for the BB code [[144,12,12]] at a physical error rate of $10^{-3}$. On the $x$-axis, only the first 10 iterations out of 400 are shown for clarity. The vertical dashed line shows the cut-off iteration number (4 iterations) to stay within the per-round time budget  of $1\,\mu s$. The figure shows that in 99.992\% of  decoding runs complete within the considered time budget. See also Appendix~\ref{app:tail-distribution-deciter} for the tail distribution of the number of decoding iterations.}
    \label{fig:convergence_etB}
\end{figure}

We also define the per-round decoding latency as the decoding latency (i.e., number of iterations multiplied by the per-iteration latency) divided by the number of syndrome measurement rounds.
 For a physical error rate of 0.001, considering the average number of decoding iterations reported in Table~\ref{tab:iterations}, the resulting per-round average decoding latency is $1.13 \times 2900 / 12 = 273$\,ns. 
Moreover, Fig.~\ref{fig:convergence_etB} shows that, at the same physical error rate, more than 99.99\% of decoding runs complete within the $1\,\mu\text{s}$ per-round time budget.

We also note that our decoder has a worst-case latency corresponding to the situation in which it reaches the maximum of 400 decoding iterations, which occurs with a probability approximately equal to the logical error rate (it is also possible for the decoder to converge on the very last iteration, but this is rare, as suggested by Fig.~\ref{fig:convergence_etB}). However, we did not attempt to optimize the maximum number of iterations, and similar accuracy could likely be achieved with a much lower maximum number of decoding iterations (Fig.~\ref{fig:convergence_etB} again provides some supporting evidence for this). Moreover, for decoders with variable latency, it has been shown in Ref.~\cite{google2025quantum} (see also the analysis in Ref.~\cite{maurya2025fpga}) that even when the worst-case latency exceeds the available time budget, the backlog problem~\cite{RevModPhys.87.307,Battistel_2023} can be avoided, provided that the decoding latency tail is not too long and occurs with sufficiently low probability.

If even lower latency is desired in an FPGA implementation, different optimizations can be explored. Firstly, the latency in our architecture is primarily limited by the $D_X$ and $D_Z$ units, due to the sequential updates imposed by the serial schedule. One possible approach to reduce this latency is to divide the $D_X$ and $D_Z$ submatrices into layers, allowing more checks to be updated in parallel without compromising accuracy. This approach requires a careful balance between the FPGA's logic and memory resources, the desired speedup, and scalability with the code distance. In addition, 
 quantization improvements could be explored to reduce the number of bits used to represent the exchanged messages in serial and layered decoders, which is currently set to 12 bits to achieve accuracy close to that of a floating-point implementation. Also, compression techniques applied to the exchanged messages~\cite{MOSTAFAPOUR2025105743} or simplifications on the check node~\cite{catala2019second,Petro2020Reduced} can be explored to reduce the delay derived from routing or logic, respectively.

Finally, we note that, using the same 16nm technology for ASIC, it is reasonable to expect, with the same design and without any further optimization, at least a three times speedup~\cite{Kuon2007Measuring}.

\heading{Discussion}
We presented a method to effectively account for correlations between Pauli errors in the detector error model, via graph augmentation and rewiring for inference. The proposed method, GARI, eliminates the 4-cycles introduced by $Y$ errors in the correlated detector error model matrix $\demxyz$, thereby shifting decoding to an augmented matrix $\demxyzbar$ designed to support MP decoding more effectively. We then identified and explored the MP decoder design choices to better leverage GARI, proposing the use of an NMS decoder with a hybrid serial-layered schedule that supports a hardware-efficient architecture.
Coupling GARI-NMS decoding with a randomized ensemble approach improved accuracy by approximately $20\times$ over BPOSD, achieving a logical error rate per round of $(6.70 \pm 1.93) \times 10^{-9}$ at a physical error rate of $10^{-3}$ for the BB code $[[144,12,12]]$ under uniform depolarizing circuit-level noise.

Finally, we investigated the FPGA implementation of our decoder and showed that, for the above BB code and physical error rate, our method achieves an average per-round decoding latency of 273\,ns and sub-microsecond latency in 99.99\% of the decoding instances. Thus, it sets a new benchmark for high-accuracy, real-time decoders and opens the way to their practical realization in fault-tolerant quantum computing systems.

It might seem that belief (message-passing) is all we need, but this would be a hasty conclusion. MP decoders are not guaranteed to work at arbitrarily large distances and may exhibit error floors at low physical/logical error rates. However, our results provide strong evidence that they can perform effectively at practically relevant distances. Moreover, a positive outcome from our simulations is that most decoding failures are due to non-convergence of the NMS decoder (i.e., the decoder failing to find an error that satisfies the syndrome), rather than incorrect error predictions. This may help eliminate error floors at lower error rates (if any) through post-processing. Indeed, in cases of non-convergence, post-processing could improve accuracy without significantly increasing the average decoding latency, since it would only be required once in a million or billion decoding samples.

We also believe that the proposed GARI approach can serve as a framework accommodating further optimizations and trade-offs, some of which have been discussed throughout the paper, and which we summarize and discuss further below. 

\emph{Decoding graph optimizations:} As discussed in \natcomm{the main text}\arxiv{Section~\ref{sec:graph}}, the GARI transform can be further applied to eliminate $4$-cycles from $\demx$ and $\demz$ matrices. This optimization, not pursued in this work, could further improve the convergence and accuracy of MP decoders and may become important when decoding larger distances, as the number of 4-cycles in $\demx$ and $\demz$ increases with the code distance.

\emph{MP decoding optimizations:} We may also consider more sophisticated MP-based decoders operating on the GARI matrix $\demxyzbar$, such as memory-BP~\cite{kuo2022exploitingdegeneracyNature,chen2025improved}, Relay-BP~\cite{muller2025improvedbeliefpropagationsufficient}, machine-learned MP~\cite{maan2025machine}. Decoders that use an MP decoder as a first stage, such as~\cite{criger2018multi,higgott2023improved,wu2025minimum,caune2023belief}, could also benefit from the GARI framework. 

\emph{Ensemble decoding optimizations:} Our ensemble decoding approach is based on the randomization of the serial schedule. Alternatively, ensemble decoding approaches based on multiple message-update rules can also be considered, e.g., through the use of randomized normalization factors or NS-FAID decoders~\cite{nguyen2017analysis}. 

\emph{Hardware and architectural optimizations:}  To further reduce latency, layered decoding can also be applied to $\demx$ and $\demz$ matrices, which first requires a careful search for a suitable layer decomposition\arxiv{ (see Section~\ref{sec:decoding-latency})}. The layered decoding can then be combined with random scheduling of layers~\cite{ducrest2023layered}, instead of one row at a time. This may also help balance the decoding latency between the top ($\demx$, $\demz$) and the bottom parts of $\demxyzbar$, which could be further exploited through pipelining and resource sharing as discussed in \natcomm{the text}\arxiv{Section~\ref{sec:hw-architecture}}.

\medskip  Finally, we note recent advances in quantum computing with neutral atoms~\natcomm{\cite{bluvstein2024logical}}\arxiv{\cite{bluvstein2024logical,manetsch2025tweezer}} 
-- particularly the implementation of transversal gates enabling low-overhead fault tolerance~\cite{zhou2024algorithmic} 
and the development of fast correlated decoding protocols~\natcomm{\cite{cain2024correlated}}\arxiv{\cite{cain2024correlated,cain2025fast,serra2025decoding}} 
-- which have shown considerable promise. However, the application of efficient MP decoders to address correlated errors in this context remains largely unexplored. We believe that the GARI framework can be adapted to effectively address such correlated errors in logical quantum computations.

\arxiv{\appendix \section*{Appendix}}
\natcomm{\heading{Methods}}

\subheading{Message Update Schedules}
\arxiv{\label{app:mp-schedules}}
Besides the update rules used to compute the exchanged messages, a key component of the MP decoder is the message update schedule, which specifies the order in which exchanged messages are updated and passed to their neighbors.  While the time complexity of a single MP decoding iteration is linear in the size of the decoding graph, regardless of the message update schedule, the choice of the schedule can affect the parallelization of message updates and the convergence speed, thus playing an important role in the achievable latency.

In the \emph{flooding (or parallel) schedule}, all nodes of the same type update their outgoing messages in parallel, alternating between error nodes and check nodes at each demi-iteration. 
The flooding schedule matches fully parallel implementations, enabling high speed but requiring substantial hardware resources and potentially facing routing and interconnect congestion issues~\cite{boutillon2014hardware}. 

In the \emph{serial schedule}, message updates are performed sequentially over a given type of nodes. The serial schedule can be \emph{vertical}, when updates are sequentialized over error nodes (i.e., column-wise), or \emph{horizontal}, when updates are sequentialized over check nodes (i.e., row-wise). The serial schedule corresponds to a fully serial implementation, minimizing hardware resources but increasing decoding latency. 

The \emph{layered schedule} updates messages layer by layer (see below), combining parallelism within each layer with sequential updates across layers. Depending on the vertical or horizontal variant, a layer consists of a set of error nodes (columns) or check nodes (rows) that do not share common neighbors (i.e., are non-intersecting), enabling efficient parallelization of message updates within the layer. By combining parallelism and sequential updates, the layered schedule is the natural choice for a partly parallel implementation, offering a balance between speed and hardware complexity~\cite{boutillon2014hardware}.

For classical LDPC codes, the flooding schedule generally achieves the same accuracy as the serial and layered schedules but requires roughly twice as many iterations to converge~\cite{zhang2007iterative}. For QLDPC codes, however, the flooding schedule may fall well short of the decoding accuracy of the serial or layered schedules, even with significantly more iterations~\cite{ducrest2022stabilizer, ducrest2023layered}. Additionally, serial and layered schedules may be randomized, further improving the decoding performance~\cite{ducrest2023layered, koutsioumpas2025colour}. In a randomized serial (or layered) schedule,  message updates are performed sequentially over the nodes (or layers) in an order randomly chosen at each decoding iteration. The asymmetry introduced by such randomization may help the decoder break the symmetry or cyclic dependencies that trap it in persistent or repetitive error patterns, thus improving convergence and overall decoding performance.

\begin{figure}[!t]
    \centering
    \includegraphics[width=\linewidth]{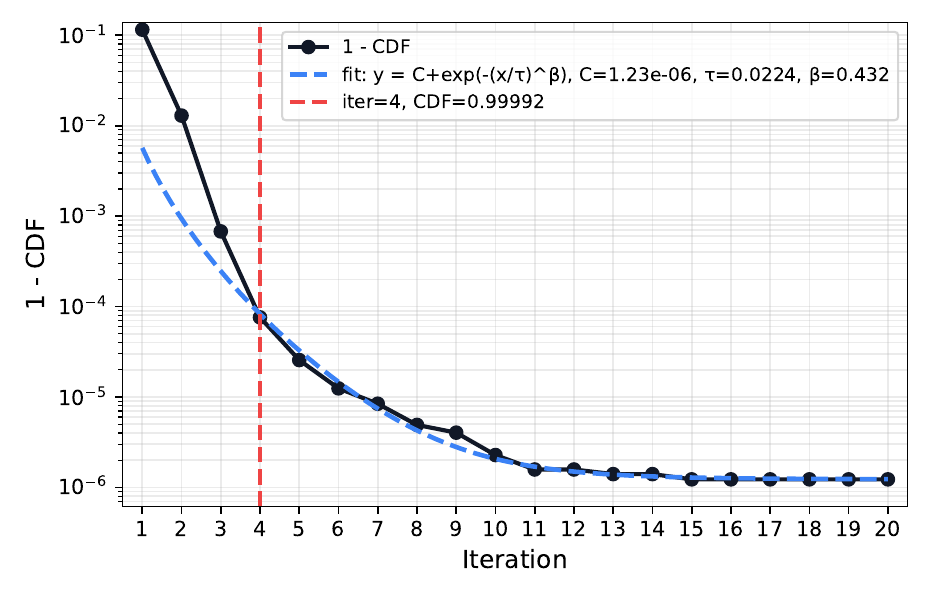}
    \caption{Tail distribution ($1-\text{CDF}$) of the number of decoding iterations for the BB code [[144,12,12]] at a physical error rate of $10^{-3}$.  On the $x$-axis, only the first 20 iterations out of 400 are shown for clarity. The vertical dashed line shows the cut-off iteration number (4 iterations) to stay within the per-round time budget  of $1\,\mu s$. The figure shows that the tail distribution decays exponentially, plateaus at around iteration 10 and has a long tail, thus showing that running more iterations does not improve the convergence further. Moreover, already at iteration 4, 99.992\% of decoding runs are completed to stay within the considered time budget.
    }
    \label{fig:convergence_fit}
\end{figure}

\subheading{Tail Distribution of the Number of Decoding Iterations}
\label{app:tail-distribution-deciter}
Fig.~\ref{fig:convergence_fit} shows the tail distribution function (i.e.,  the complementary cumulative distribution function, $1-\text{CDF}$, ) of the number of decoding iterations for the BB code [[144,12,12]] at a physical error rate of $10^{-3}$.  
We illustrate the tail distribution up to 20 decoding iterations. 

\subheading{BPOSD Tuning}
\arxiv{\label{app:BPOSD_sr_vs_pl}}

Here, we discuss simulation results for the order-zero BPOSD operating on the $\demz$ and $\demxyz$ matrices.

\begin{figure*}
    \centering
    \begin{subfigure}[b]{0.45\textwidth}
    \captionsetup{format=centercaption}
    \setlength{\abovecaptionskip}{0pt}
        \centering
        \includegraphics[width=\textwidth]{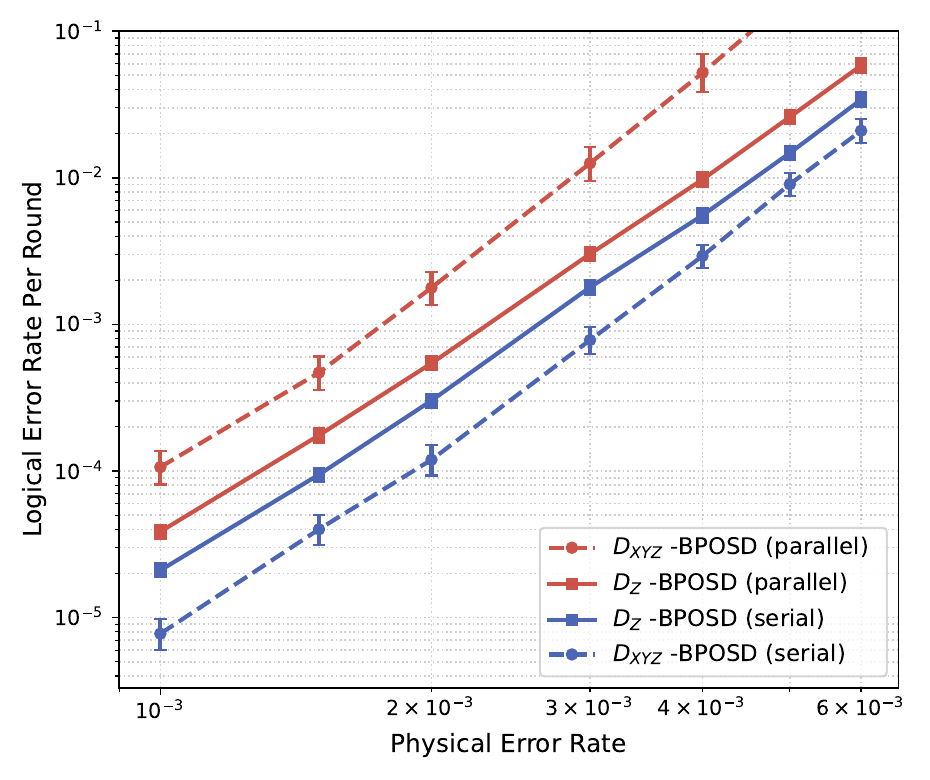}
        \caption{[[72,12,6]] code}
        \label{fig:BPOSD_sr_vs_pl-d6}
    \end{subfigure}
    \hfill
    \begin{subfigure}[b]{0.45\textwidth}
    \captionsetup{format=centercaption}
    \setlength{\abovecaptionskip}{0pt}
        \centering
        \includegraphics[width=\textwidth]{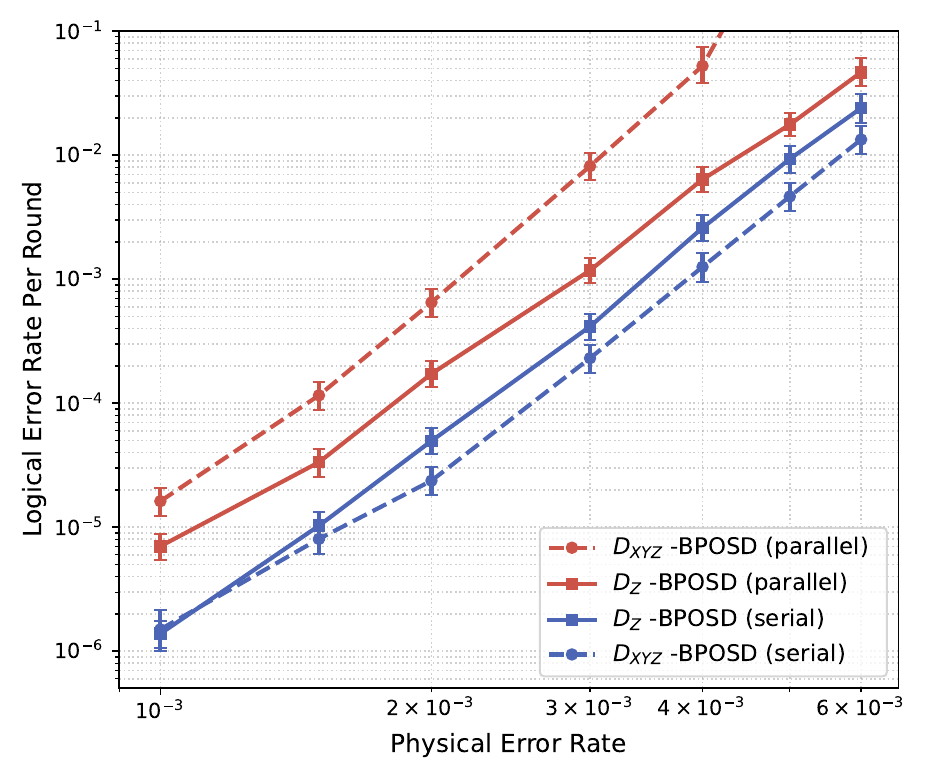}
        \caption{[[90,8,10]] code}
        \label{fig:BPOSD_sr_vs_pl-d10}
    \end{subfigure}

    \vskip 3mm  

    \begin{subfigure}[b]{0.45\textwidth}
    \captionsetup{format=centercaption}
    \setlength{\abovecaptionskip}{0pt}
        \centering
        \includegraphics[width=\textwidth]{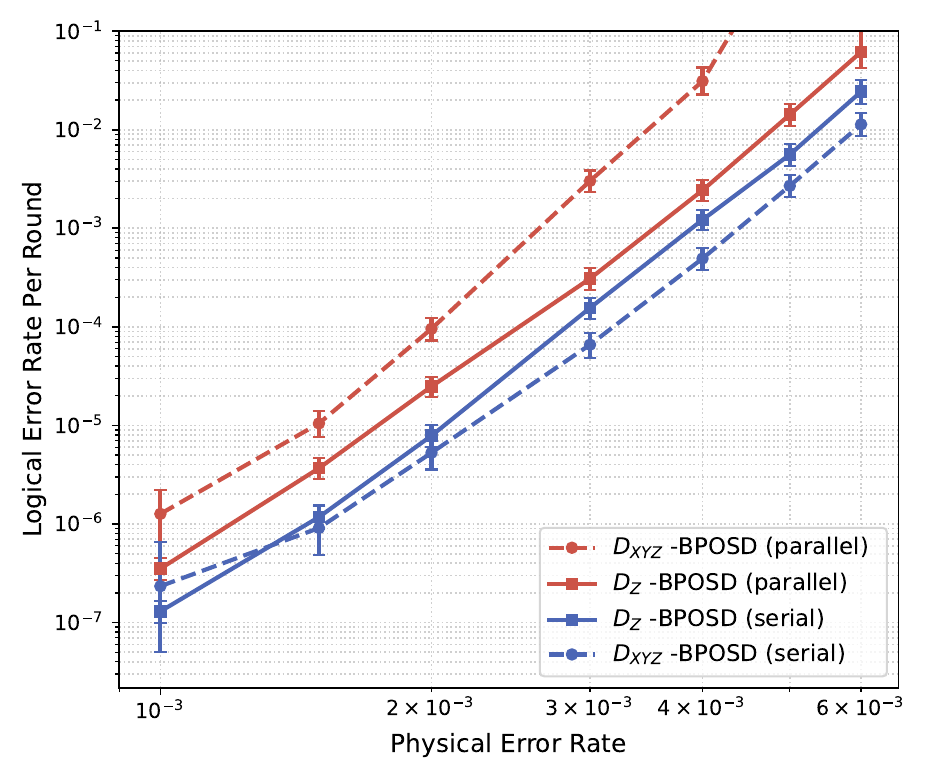}
        \caption{[[144,12,12]] code}
        \label{fig:BPOSD_sr_vs_pl-d12}
    \end{subfigure}
    \caption{Logical error rate performance of BPOSD decoder running on $\demz$ and $\demxyz$ matrices, with either parallel or serial schedule, for BB codes under uniform depolarizing circuit level noise. Error bars indicate 99\% confidence intervals. See discussion in the text.
}
    \label{fig:BPOSD_sr_vs_pl}
\end{figure*}

For the distance-6 BB code, when the parallel (flooded) scheduling is used for the underlying MP decoder, it was observed in~\cite[Appendix~A]{beni2025tesseractsearchbaseddecoderquantum} that \emph{uncorrelated BPOSD is significantly more accurate than correlated BPOSD}, that is BPOSD performs better when operating on $\demz$ than  $\demxyz$. We reproduce these results in Fig.~\ref{fig:BPOSD_sr_vs_pl}(a), and provide similar results for BB codes with distances 10 and 12 in Fig.~\ref{fig:BPOSD_sr_vs_pl}(b)-(c) -- see orange curves, corresponding to parallel scheduling.  

Additionally, we perform simulations using a randomized serial scheduling (blue curves), and observe a significant improvement in the BPOSD performance, for both $\demz$ and $\demxyz$ matrices.

Interestingly, for the distance-6 BB code, BPOSD with randomized serial scheduling performs significantly better on $\demxyz$  than on $\demz$.  Thus,  accuracy results included in the main text for comparison purposes correspond to the dashed blue curve in Fig.~\ref{fig:BPOSD_sr_vs_pl}(a).

For BB codes with distances 10 and 12, BPOSD with serial scheduling outperforms parallel scheduling at higher physical error rates. However, the performance gap diminishes, and may even reverse, at lower physical error rates, as shown in Fig.~\ref{fig:BPOSD_sr_vs_pl}(b) and (c). As BPOSD with randomized serial scheduling on $\demz$ achieves the best performance and exhibits a steeper slope in the low-error-rate regime, we adopt this variant of the decoder as a baseline for comparison with our method in \natcomm{the main text}\arxiv{Section~\ref{sec:accuracy}}, for BB codes with distances 10 and 12.

Finally, we note that all simulation results in Fig.~\ref{fig:BPOSD_sr_vs_pl} use an NMS decoder, with the normalization factor optimized for the best performance under the considered scheduling. Specifically, we use a normalization factor of 1.0 for parallel scheduling and 0.625 for serial scheduling. The maximum number of decoding iterations is set to 30, as we observed that a larger number of iterations degrades performance at low physical error rates due to the erratic behavior of the NMS decoder in case of non-convergence.

\section*{Data Availability}
The simulation data generated in this study has been deposited in the Zenodo database and is available at~\cite{maan_2025_17991975}.

\section*{Code Availability}
The code used to generate the GARI matrices and perform the numerical simulations in this paper is available at~\cite{github_gari_nms}

\natcomm{\bibliography{__main}}

\section*{Acknowledgments}
We thank Tristan Müller for helpful discussion regarding simulations of the Relay-BP decoder. We also thank Laleh Aghababaie Beni, Noah Shutty and Oscar Higgott for helpful discussions and sharing Tesseract decoder data with us. We thank Timo Hillmann for the discussion on the challenges of decoding correlated errors using BPOSD. 

A. S. Maan and A. Paler acknowledge
funding from the Defense Advanced Research Projects Agency [under the Quantum Benchmarking (QB) program, contracts no. HR00112230006 and HR001121S0026], and the QuantERA grant EQUIP through the Academy of Finland, decision number 352188. The views, opinions and/or findings expressed are those of the author(s) and should not be interpreted as representing the official views or policies of the Department of Defense or the U.S. Government.

F. Garcia-Herrero acknowledges support from the project PID2023-147059OB-I00 funded by MCIU/ AEI/ 10.13039/501100011033/ FEDER.  UE. His work was also partially funded by a grant from Google Quantum AI. 

V. Savin acknowledges support from the Plan France 2030, project no. ANR-22-PETQ-0006  (NISQ2LSQ) and Programme d’investissement d’avenir, IRT Nanoelec, project no. ANR-10-AIRT-05 (Q-Loop).

The calculations presented above were performed using computer resources within the Aalto University School of Science “Science-IT” project. We acknowledge the EuroHPC Joint Undertaking for awarding this project access to the EuroHPC supercomputer LUMI, hosted by CSC (Finland) and the LUMI consortium through a EuroHPC Regular Access call.

\section*{Author Contributions}
ASM and AP observed that serial scheduling improves BPOSD performance for the correlated detector error model. VS, ASM and FMGH collaborated on the development of GARI decoding during an EQUIP meeting in Madrid, where ASM highlighted the regular structure of the decoding matrices, VS contributed the GARI decoding formulation, and FMGH noted the FPGA compatibility of the approach. FMGH introduced the ensemble decoding and the layered decoding methods, while ASM, VS and AP refined the ensemble method. The software implementation of the decoder was developed by ASM and FMGH, and the hardware implementation by FMGH. ASM tuned the decoder and ran the numerical simulations. VS was responsible for the writing of the manuscript, and all other authors provided input and revisions. AP and VS verified the results.

\section*{Competing Interests}
The authors declare no competing interests.

\arxiv{\bibliography{__main}}

\end{document}

%% file: figs/tikz/4cycle.tex

\tikzset{
    vnode/.style={draw, circle, minimum size=6mm, thick},
    cnode/.style={draw, rectangle, minimum size=6mm, thick},
    edge/.style={thick}
}

\begin{tikzpicture}[x=1cm, y=1cm, every node/.style={scale=1}]
    \node[cnode] (c1) at (0,0) {};
    \node[cnode] (c2) at (4,0) {};

    \node[vnode] (v1) at (0,3) {};
    \node[vnode] (v2) at (4,3) {};

    \draw[edge,thick] (v1.south) -- (c1.north);
    \draw[edge,thick] (v1.south) -- (c2.north);
    \draw[edge,thick] (v2.south) -- (c1.north);
    \draw[edge,thick] (v2.south) -- (c2.north);
    
    \draw[dashed] (v1.south) -- ($(v1.south) - (1.5,0.75)$);
    \draw[dashed] (v1.south) -- ($(v1.south) + (0.5,-0.75)$);
    \draw[dashed] (v2.south) -- ($(v2.south) + (1.5,-0.75)$);
    \draw[dashed] (v2.south) -- ($(v2.south) - (0.5,0.75)$);
    
    \draw[dashed] (c1.north) -- ($(c1.north) + (-1.75,0.75)$);
    \draw[dashed] (c1.north) -- ($(c1.north) + (-0.8, 0.75)$);
    \draw[dashed] (c1.north) -- ($(c1.north) + (0.5,0.75)$);
    
    \draw[dashed] (c2.north) -- ($(c2.north) + (1.75,0.75)$);
    \draw[dashed] (c2.north) -- ($(c2.north) + (0.8, 0.75)$);
    \draw[dashed] (c2.north) -- ($(c2.north) + (-0.5,0.75)$);
    
\end{tikzpicture}

%% file: figs/tikz/biclique.tex

\tikzset{
    vnode/.style={draw, circle, minimum size=6mm, thick},
    cnode/.style={draw, rectangle, minimum size=6mm, thick},
    edge/.style={thick}
}

\begin{tikzpicture}[x=1cm, y=1cm, every node/.style={scale=1}]
    \node[cnode] (c1) at (0,0) {};
    \node[cnode] (c2) at (2,0) {};
    \node[cnode] (c3) at (4,0) {};
    \node[cnode] (c4) at (6,0) {};

    \node[vnode] (v1) at (1,3) {};
    \node[vnode] (v2) at (3,3) {};
    \node[vnode] (v3) at (5,3) {};

    \node[draw=none, inner sep=0, font=\large] at (v1) {$i$};
    \node[draw=none, inner sep=0, font=\large] at (v2) {$j_1$};
    \node[draw=none, inner sep=0, font=\large] at (v3) {$j_2$};

    \draw[edge,thick] (v1.south) -- (c1.north);
    \draw[edge,thick] (v1.south) -- (c2.north);
    \draw[edge,thick] (v1.south) -- (c3.north);
    \draw[edge,thick] (v1.south) -- (c4.north);
    
    \draw[edge,thick] (v2.south) -- (c1.north);
    \draw[edge,thick] (v2.south) -- (c2.north);
    \draw[edge,thick] (v2.south) -- (c3.north);
    \draw[edge,thick] (v2.south) -- (c4.north);
    
    \draw[edge,thick] (v3.south) -- (c1.north);
    \draw[edge,thick] (v3.south) -- (c2.north);
    \draw[edge,thick] (v3.south) -- (c3.north);
    \draw[edge,thick] (v3.south) -- (c4.north);
    
\end{tikzpicture}

%% file: figs/tikz/gari.tex

\tikzset{
    vnode/.style={draw, circle, minimum size=6mm, thick},
    cnode/.style={draw, rectangle, minimum size=6mm, thick},
    vpnode/.style={draw, circle, minimum size=6mm, fill=blue, thick},
    cpnode/.style={draw, rectangle, minimum size=6mm, fill=blue, thick},
    edge/.style={thick}
}

\begin{tikzpicture}[x=1cm, y=1cm, every node/.style={scale=1}]
    \node[cnode] (c1) at (0,0) {};
    \node[cnode] (c2) at (2,0) {};
    \node[cnode] (c3) at (4,0) {};
    \node[cnode] (c4) at (6,0) {};
    \node[cpnode] (cp) at (8,0) {};

    \node[vnode] (v1) at (1,3) {};
    \node[vnode] (v2) at (3,3) {};
    \node[vnode] (v3) at (5,3) {};
    \node[vpnode] (vp) at (7,3) {};

    \node[draw=none, inner sep=0, font=\large] at (v1) {$i$};
    \node[draw=none, inner sep=0, font=\large] at (v2) {$j_1$};
    \node[draw=none, inner sep=0, font=\large] at (v3) {$j_2$};

    \node[draw=none, inner sep=0, text=white, font=\large] at (vp) {$i$};
    \node[draw=none, inner sep=0, text=white, font=\large] at (cp) {$i$};

    \draw[edge,thick] (v1.south) -- (cp.north);
    \draw[edge,thick] (v2.south) -- (cp.north);
    \draw[edge,thick] (v3.south) -- (cp.north);
    \draw[edge,thick] (vp.south) -- (cp.north);
    
    \draw[edge,thick] (vp.south) -- (c1.north);
    \draw[edge,thick] (vp.south) -- (c2.north);
    \draw[edge,thick] (vp.south) -- (c3.north);
    \draw[edge,thick] (vp.south) -- (c4.north);
    
\end{tikzpicture}

%% file: figs/tikz/hw_archi_v3.tex
\begin{tikzpicture}[x=1cm, y=1cm, every node/.style={scale=1}]

\pgfmathsetlengthmacro{\hsp}{0.6cm}  
\pgfmathsetlengthmacro{\vsp}{1.2cm}

\pgfmathsetlengthmacro{\DXwidth}{2.2cm}  
\pgfmathsetlengthmacro{\DXheight}{2cm}
\pgfmathsetlengthmacro{\DZwidth}{2.2cm}  
\pgfmathsetlengthmacro{\DZheight}{2.5cm}
\pgfmathsetlengthmacro{\Bwidth}{9cm}  
\pgfmathsetlengthmacro{\Bheight}{3cm}

\tikzset{
  DXunit/.style={
     draw, rectangle,
     minimum width=\DXwidth,
     minimum height=\DXheight,
     inner sep=0pt,
     align=center
  },
  DZunit/.style={
     draw, rectangle,
     minimum width=\DZwidth,
     minimum height=\DZheight,
     inner sep=0pt,
     align=center
  },
  Bunit/.style={
     draw, rectangle,
     minimum width=\Bwidth,
     minimum height=\Bheight,
     inner sep=0pt,
     align=center
  }
}

\pgfmathsetlengthmacro{\hsp}{0.6cm}  
\pgfmathsetlengthmacro{\vsp}{1.2cm}

\node[DZunit] (DZ) {};
\node[text width=\DZwidth, anchor=north, align=center] at ([yshift=-0.32*\vsp] DZ.north) {$D_Z$ Unit\\ {\small (serial sched.)}};
\node[rectangle, draw=black, dashed, inner sep=2pt, text width=1.7cm, anchor=north, , align=center, font=\small] at ([yshift=-1.27*\vsp] DZ.north) (es) {$D_Z \bar{e}_X = s_Z$\\[-1mm] or max-iter};

\node[DXunit, anchor=north east] (DX) at ([xshift=-\hsp] DZ.north west) {};
\node[text width=\DXwidth, anchor=north, align=center] at ([yshift=-0.32*\vsp] DX.north) {$D_X$ Unit\\ {\small (serial sched.)}};

\node[Bunit, anchor=north east, draw=none] (BU) at ([yshift=-\vsp] DZ.south east) {};

\draw[->] (DX.south) -- (DX.south |- BU.north);
\draw[<-] (DX.north) -- ++(0,\vsp/3) -- ++(-\DXwidth/2-\hsp/2,0) coordinate (temp) -- (temp |- BU.north);
\draw[dashed]  (temp |- BU.north) -- (temp |- BU.south);
\draw[-] (temp |- BU.south) -- ++(0,-\vsp/3) -- ++(\DXwidth/2+\hsp/2, 0) -- ++(0,\vsp/3); 

\node[text width=1.3cm, inner xsep=1pt, inner ysep=5pt, align=center, anchor=north west, font=\small] at (temp |- DZ.south) {extrinsic\\[-1mm] messages\\[-1mm] for $\bar{e}_Z$};

\draw[->] (DZ.south) -- (DZ.south |- BU.north);
\draw[<-] (DZ.north) -- ++(0,\vsp/3) -- ++(-\DZwidth/2-\hsp/2,0) coordinate (temp) -- (temp |- BU.north);
\draw[dashed]  (temp |- BU.north) -- (temp |- BU.south);
\draw[-] (temp |- BU.south) -- ++(0,-\vsp/3) -- ++(\DZwidth/2+\hsp/2, 0) -- ++(0,\vsp/3); 

\node[text width=1.3cm, inner xsep=1pt, inner ysep=5pt, align=center, anchor=north west, font=\small] at (temp |- DZ.south) {extrinsic\\[-1mm] messages\\[-1mm] for $\bar{e}_X$};

\node[Bunit, anchor=north east, fill=white, fill opacity=0.7] at ([yshift=-\vsp] DZ.south east) {};
\node[text width=3cm, align=center] at (BU.center) {Bottom Part Unit\\ {\small (2-layers schedule)}};

\coordinate (temp) at ($(BU.north west)!0.5!(BU.north)$);
\draw[<-] (temp) -- ++(0, \vsp/2) -- ++(-\Bwidth/3.5, 0);
\node[text width=2.6cm, align=center, font=\small, anchor=south east] at ([yshift=3pt]temp) {prior probabilities\\ for $(e_Z, e_X, e_Y)$};

\coordinate (temp) at ($(DX.north east)!0.5!(DX.north)$);
\draw[<-] (temp) -- ++(0,\vsp/3) coordinate (temp);
\node[align=center, font=\small, anchor=south] at (temp) {$s_X$};

\coordinate (temp) at ($(DZ.north east)!0.5!(DZ.north)$);
\draw[<-] (temp) -- ++(0,\vsp/3) coordinate (temp);
\node[align=center, font=\small, anchor=south] at (temp) {$s_Z$};

\coordinate (temp) at ($(es.south east)!0.5!(es.south)$);
\draw[->] (temp) -- ++(0,-\vsp/2) -- ++(\hsp,0) node[anchor=west, font=\small] (bareX) {$\bar{e}_X$};
\node[below right=-0.15cm and -1.4cm of bareX, font=\small, text width=1.5cm, align=center] {estimated\\[-1mm] error};

\end{tikzpicture}